\documentclass[aps,pre,reprint,superscriptaddress,amsmath,showpacs,amssymb]{revtex4-2}

\usepackage[T1]{fontenc}

\usepackage{empheq}

\usepackage{booktabs}

\usepackage{epsfig}
\usepackage{graphicx}
\usepackage{bm}
\usepackage{color}
\usepackage[hyperindex,breaklinks,unicode,colorlinks=true,linkcolor=blue,citecolor=blue,urlcolor=blue]{hyperref}

\usepackage[draft]{changes}

\usepackage{tikz}
\usepackage{tikz-3dplot}
\usepackage{tkz-euclide}
\usetikzlibrary{decorations.markings}

\usepackage{multirow}
\usepackage{makecell}

\usepackage{colortbl}
\usepackage{xcolor}
\usepackage{array}

\usepackage{mathrsfs}

\usetikzlibrary{external}
\begin{document}

\newcommand\JXNU{\affiliation{College of Physics, Communication and Electronics, Jiangxi Normal University, Nanchang 330022, China}}

\newcommand\uleip{\affiliation{Institut f\"ur Theoretische Physik, Universit\"at Leipzig,  Postfach 100 920, D-04009 Leipzig, Germany}}

\newcommand\NCHU{\affiliation{Key Laboratory of Opto-Electronic Information Science and Technology of Jiangxi Province, Nanchang Hangkong University, Nanchang 330063, China}}

\newcommand\chau{\affiliation{ 
 Department of Macromolecular Physics,  
 Faculty of Mathematics and Physics, Charles University,
 V Hole{\v s}ovi{\v c}k{\' a}ch 2, 
 CZ-180~00~Praha, Czech Republic
}}

\title{Optimizing low-dissipation Carnot-like thermal devices with heat leak}

\author{Zhuolin Ye}\email{zhuolinye@foxmail.com}\JXNU
\author{Viktor Holubec}\email{viktor.holubec@mff.cuni.cz}\chau

\begin{abstract}
Delimiting the bounds of optimal performance for heat engines (HEs), refrigerators (REs), and heat pumps (HPs) is a central goal in thermodynamics. While low-dissipation (LD) models have proven valuable for this purpose, the role of heat leak in such models has received limited attention. Here, we present a unified framework for LD Carnot-like (CL) HEs, REs, and HPs that incorporates heat leaks, and derive new results for the efficiency at maximum power and the power at maximum efficiency. We further investigate the relationship between the bounds of power at fixed efficiency and efficiency at fixed power, demonstrating that these bounds coincide and are described by identical curves across all thermal devices. Finally, we show that the optimal performance of all three devices can be achieved by optimizing the average entropy production rate over the cycle, a result that holds for any CL device and extends beyond the LD assumption.
\end{abstract}

\maketitle


\section{Introduction}
\label{introduction}

Thermal devices (TDs) such as heat engines (HEs), refrigerators (REs), and heat pumps (HPs) are fundamental to modern technology. HEs, which convert thermal energy into mechanical work, were central to the Industrial Revolution~\cite{marks2007origins}. REs, developed in the 19th and 20th centuries, enabled reliable food preservation by transferring heat from cold regions~\cite{thévenot1979history,Pearson2022}. More recently, HPs have gained prominence for their ability to heat homes efficiently by extracting thermal energy from the environment~\cite{Pearson2022}.

A central goal in engineering is to optimize these devices to deliver sufficient power as efficiently and cost-effectively as possible. However, due to the complexity of real systems, direct analytical optimization is often intractable. To address this, physicists introduced simplified models that allow for analytical optimization~\cite{Tu_2012}, offering fundamental insights into optimal performance. Among these, low-dissipation (LD) models~\cite{PhysRevLett.105.150603} play a prominent role. They assume that the entropy produced during isothermal processes is inversely proportional to their duration, an assumption valid for slowly driven systems~\cite{PhysRevA.21.2115}, and exact in specific setups such as optimized overdamped Brownian HEs~\cite{schmiedl2007efficiency}. 

In contrast to LD models, endoreversible models~\cite{curzon1975efficiency,Chen1989TheEO,de1985efficiency} require assumptions about specific heat-transfer laws, while linear irreversible models~\cite{PhysRevLett.95.190602,PhysRevLett.112.180603,PhysRevE.77.041127} assume operation within the linear-response regime of small forces and temperature gradients. Due to their generality and analytical simplicity, LD models~\cite{PhysRevE.85.010104,PhysRevE.103.052125,PhysRevE.105.024139} have been optimized from multiple perspectives~\cite{PhysRevLett.124.110606,PhysRevE.87.012105,PhysRevE.96.042128}. 

An important factor that was initially neglected in LD models is heat leak. Earlier investigations, based on endoreversible models~\cite{moukalled1995efficiency, gordon1992general, chen2002optimal} and other studies~\cite{PhysRevE.91.050102, PhysRevE.93.042112, PhysRevE.77.041127}, have shown that the presence of heat leak significantly modifies the shape of the power-efficiency curve~\cite{chen1994maximum}. The inclusion of the heat leak term into LD models was first motivated by the physical significance of the nonlinear dissipation term in minimally nonlinear irreversible frameworks~\cite{izumida2012efficiency,PhysRevE.91.052140}. As a result, LD models with heat leak have become a paradigmatic example of non-tight coupling in minimally nonlinear irreversible models~\cite{izumida2012efficiency, PhysRevE.91.052140, PhysRevE.101.052124,PhysRevE.101.052124}. However, the role of heat leak in LD models remains insufficiently studied, and this paper aims to address this knowledge gap.

The main goals of this paper are threefold. First, in Sec.~\ref{sec:model}, we provide a unified description of LD Carnot-like (CL) HEs, REs, and HPs with heat leak. Second, in Sec.~\ref{sec:emp}, we review known results~\cite{50influence, huang2015performance} on efficiency at maximum power (EMP) for LD HEs with heat leak, and derive new results for EMP of LD REs and HPs with heat leak. In Sec.~\ref{sec:PME-fir}, we derive power at maximum efficiency (PME) for these three devices and show that they are governed by the same system of equations. In Sec.~\ref{sec:optmal-relation-fir}, we derive the bounds of power at fixed efficiency (BPFE) and the bounds of efficiency at fixed power (BEFP), showing that they are represented by identical curves in the power-efficiency diagram for arbitrary TDs. In Sec.~\ref{sec:relation-bounds-fir}, we demonstrate that the optimal performance for all three CL TDs can, in general, be achieved in a unified manner by optimizing the average entropy production rate during the cycle. Finally, we conclude in Sec.~\ref{sec-con}.

\begin{figure}[t]
\centering
\includegraphics[trim={0cm 0cm 0cm 0cm}, width=0.6\columnwidth]{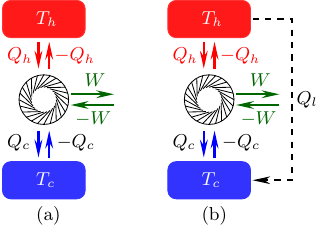}
\vspace{-0.3cm}
\caption{CL TDs: (a) without heat leak and (b) with heat leak. HEs operate in the regime where positive-sign fluxes are positive. REs and HPs operate in the regime where negative-sign fluxes are positive.
}
\label{fig:model}
\end{figure}

\section{Model}
\label{sec:model}

In Sec.~\ref{sec:generic-model}, we introduce CL HEs, REs, and HPs with heat leak in a unified manner. In Sec.~\ref{sec:LD-model-fir}, we consider the situation when these devices operate in the LD regime.

\subsection{Generic model and sign convention}
\label{sec:generic-model}

Consider a heat device operating along a finite-time forward or inverse Carnot cycle consisting of two isotherms and two adiabats, depicted in Fig.~\ref{fig:model}.
Compared to the diagrams of standard TDs in textbooks~\cite{reif2009fundamentals,bejan2016advanced,dittman2021heat}, the difference lies in the use of both positive and negative signs preceding the symbols $\{Q_h, Q_c, W\}$. This approach serves to distinguish between the forward (operating as HEs) and the inverse (operating as REs or HPs) CL cycles, allowing for a unified description of HEs, REs, and HPs even when considering the heat leak.

Specifically, for HEs, heat $Q_{ h}$ is absorbed by the system from the hot bath at temperature $T_h$ during the hot isotherm, some of which ($Q_{ c}$) is dumped into the cold bath at temperature $T_c$ during the cold isotherm, thereby producing the output work $W$. Conversely, in the case of REs and HPs, by consuming the input work $-W$,
heat $-Q_{ c}$ is extracted by the system from the cold bath ($T_c$) during the cold isotherm and $-Q_{ h}$ is evacuated to the hot bath ($T_h$) during the hot isotherm. In the presence of heat leak $Q_l$, which characterizes the direct heat transfer between the reservoirs,
the net heats released by the hot bath and injected into the cold bath per forward CL cycle are given by~\cite{chen1997influence,chen1994maximum} 
\begin{align}
\varPhi_h&=Q_h+Q_{ l}, 
\label{model-leaky-qh}\\
\varPhi_c&=Q_c+Q_l. 
\label{model-leaky-qc}
\end{align}
Nevertheless, according to the first law, $Q_l$ does not affect the output work, $W=\varPhi_h-\varPhi_c=Q_{ h}-Q_{ c}$. For the inverse CL cycle, the net heats evacuated to the hot bath and extracted from the cold bath are $-\varPhi_h$ and $-\varPhi_c$, respectively.

The total entropy production for both the forward and inverse CL cycles reads
    \begin{equation}
\Delta S_{ tot}=\frac{\varPhi_c}{T_c}-\frac{\varPhi_h}{T_h}=\left(\frac{Q_c}{T_c}-\frac{Q_h}{T_h}\right)+\left(\frac{Q_l}{T_c}-\frac{Q_l}{T_h}\right).
\label{entropy-production}
\end{equation}
Since the system operates cyclically, $\Delta S_{tot}$ represents the increase in entropy in the environment per cycle of operation of the TDs. The first term in the bracket corresponds to dissipation arising from the finite-time heat transfer between the system and the reservoirs, which vanishes in the quasistatic limit. The second term results from the direct heat transfer between the reservoirs and diverges in the quasistatic limit unless the so-called tight-coupling condition (no heat leak) is satisfied~\cite{izumida2013coefficient, izumida2012efficiency, PhysRevE.91.052140}.

The usual thermodynamic quantities used to assess the performance of TDs are their power and efficiency. 
For REs and HPs, the powers are referred to as cooling power and heating load, respectively, and their efficiencies are both termed as coefficient of performance~\cite{guo2019thermally,ahmadi2015thermo}. For convenience, we adhere to using the names `power' and `efficiency' for REs and HPs throughout this paper.
Denoting as $t_p$ the cycle period, power and efficiency for the considered CL HEs, REs, and HPs with heat leak are defined as~\cite{PhysRevE.103.052125,PhysRevE.105.024139,PhysRevE.101.052124,PhysRevE.96.062107}
\begin{empheq}[left=\text{HE}\empheqlbrace]{align}  
 P&=\frac{W}{t_{ p}}=\frac{\varPhi_h-\varPhi_c}{t_{ p}}, \label{eq:HE-power} \\  
 \eta&=\frac{W}{\varPhi_h}=1-\frac{\varPhi_c}{\varPhi_h}=\frac{\eta_C}{1+\frac{T_c\Delta S_{tot}}{Pt_p}},\quad\quad \label{eq:HE-eff} 
\end{empheq} 
\begin{empheq}[left=\text{RE}\empheqlbrace]{align}  
 R&=\frac{-\varPhi_{ c}}{t_{ p}}, \label{eq:RE-power} \\  
 \varepsilon&=\frac{-\varPhi_{ c}}{-W}=\frac{\varPhi_{ c}}{\varPhi_h-\varPhi_c}=\frac{\varepsilon_C}{1+\frac{\varepsilon_CT_h\Delta S_{tot}}{Rt_p}}, \label{eq:RE-eff}  
\end{empheq}
\begin{empheq}[left=\text{HP}\empheqlbrace]{align}  
 M&=\frac{-\varPhi_{ h}}{t_{ p}}, \label{eq:HP-power} \\  
 \epsilon&=\frac{-\varPhi_{ h}}{-W}=\frac{\varPhi_{ h}}{\varPhi_h-\varPhi_c}=\frac{\epsilon_C}{1+\frac{\epsilon_CT_c\Delta S_{tot}}{Mt_p}}. \label{eq:HP-eff}  
\end{empheq}
When the individual devices operate within their respective regimes, as shown in Fig.~\ref{fig:model}, the maximum values of $\eta$, $\varepsilon$, and $\epsilon$—which are attainable only in the reversible limit ($\Delta S_{tot} = 0$)—are given by $\eta_C = 1 - \tau$ ($\tau \equiv T_c / T_h$), $\varepsilon_C = \tau / (1 - \tau)$, and $\epsilon_C = 1 / (1 - \tau)$. The minimum value of $\eta$ is 0, reached when $\varPhi_h = \varPhi_c$ (equivalently, $Q_h = Q_c$). In this case, the heat $Q_h$ extracted from the hot bath is entirely dissipated to the cold bath (no output work), acting as heat leak. Consequently, the HEs behave as pure heat exchangers, allowing heat to spontaneously flow from the hot bath to the cold bath. Conversely, the minimum values of $\varepsilon$ and $\epsilon$ are 0 and 1, respectively, occurring when $-W = -\varPhi_h$. In this scenario, a portion ($-Q_l$) of the input work is used to counteract the heat leak, while the remainder ($-Q_h$) participates in the direct conversion of work into heat. Therefore, the REs and HPs act as work-to-heat converters, such as electric heaters converting electrical work into heat.

\subsection{LD models with dimensionless variables}
\label{sec:LD-model-fir}

Let us now shift our focus to LD CL TDs with heat leak. LD HEs and REs with heat leak were first studied in Refs.~\cite{izumida2013coefficient,PhysRevE.91.052140}. However, to the best of our knowledge, no studies have been devoted to LD HPs with heat leak.

\begin{figure}[t]
\centering
\includegraphics[trim={0cm 0cm 0cm 0cm}, width=0.85\columnwidth]{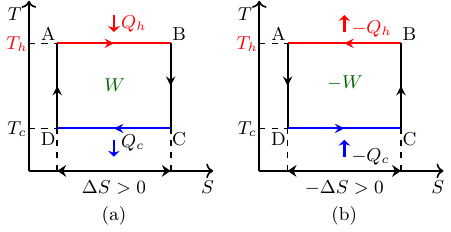}
\vspace{-0.3cm}
\caption{Bath temperature-system entropy ($T-S$) diagrams of the (a) forward (functioning as HEs) and (b) inverse (functioning as REs or HPs) CL cycles operating in the LD regime. The horizontal (vertical) lines represent isotherms (adiabats). The areas enclosed by the rectangles equal the respective works only when the cycles are executed quasistatically.
}
\label{fig:TS}
\end{figure}

For LD HEs, the transferred heats between the reservoirs and the system are expressed as~\cite{PhysRevLett.105.150603,PhysRevLett.124.110606} 
\begin{align}
Q_h&=T_h\left(\Delta S-\frac{\sigma_h}{t_h}\right), 
\label{model-he-qh}\\
Q_c&=T_c\left(\Delta S+\frac{\sigma_c}{t_c}\right), 
\label{model-he-qc}
\end{align}
where the irreversibility parameters, $\sigma_i\ge 0,~i=h, c$, quantify the extent of irreversibility in the individual isotherms, each lasting for durations $t_i$. These parameters depend on the characteristics of the system~\cite{schmiedl2007efficiency,iyyappan2020efficiency,PhysRevE.92.032113,zhao2022microscopic} and can be easily measured~\cite{martinez2016brownian}. It is observed from Fig.~\ref{fig:TS}(a) that $\Delta S>0$ denotes the increase (decrease) of the system entropy during the hot (cold) isotherm for the forward CL cycle. Based on Eqs.~\eqref{model-he-qh} and~\eqref{model-he-qc}, the transferred heats between the reservoirs and the system for LD REs and HPs can be written as
\begin{align}
-Q_h&=T_h\left(-\Delta S+\frac{\sigma_h}{t_h}\right), 
\label{model-he-qh-mod}\\
-Q_c&=T_c\left(-\Delta S-\frac{\sigma_c}{t_c}\right). 
\label{model-he-qc-mod}
\end{align}
From Fig.~\ref{fig:TS}(b) one finds that $-\Delta S>0$ stands for the decrease (increase) of the system entropy during the hot (cold) isotherm for the inverse CL cycle. Altogether, $\Delta S$ can be employed to represent the increase (decrease) of the system entropy during the hot (cold) isotherm for both the forward ($\Delta S>0$) and the inverse ($\Delta S<0$) CL cycles.

According to Newton's law of cooling~\cite{chen1994maximum,PhysRevE.97.022139}, the heat leak between the reservoirs can be formulated as
\begin{equation}
Q_l=\kappa t_{ p}(T_{ h}-T_{ c}),
\label{example-fir-exp-ql}
\end{equation}
where $\kappa$ is the thermal conductivity. 
In the following, we employ the assumption, widely used in theoretical works~\cite{schmiedl2007efficiency,dechant2017underdamped,ye2022optimal,curzon1975efficiency,yan1990class,PhysRevLett.105.150603,PhysRevE.88.062115,holubec2016maximum,PhysRevE.101.052124,PhysRevE.105.024139,PhysRevE.103.052125} and realized in experiments~\cite{blickle2012realization}, that the adiabats are instantaneous. Accordingly, the cycle period can be approximated as $t_p\approx t_h+t_c$.

Before inserting Eqs.~\eqref{model-he-qh}-\eqref{example-fir-exp-ql} into Eqs.~\eqref{model-leaky-qh}-\eqref{entropy-production}, we first introduce four dimensionless variables:
\begin{equation}
\left(
\begin{array}{cc}
\tilde{\kappa}  &  \tilde{t}_p\\
\alpha       & \sigma
\end{array} 
\right)=
\left(
\begin{array}{cc}
 \kappa\sigma_h/\Delta S^2 & t_p\Delta S/\sigma_h \\
t_h/t_p       & \sigma_c/\sigma_h
\end{array} 
\right),
\label{append-reduced-matrix}
\end{equation}
where $\tilde{\kappa}$ is the reduced thermal conductivity, $\tilde{t}_p$ is the reduced cycle period, $\alpha\in[0, 1]$ is the reduced contact time reflecting the allocation of the cycle period between the two isotherms of the cycle, and $\sigma$ is the so-called irreversibility ratio studied in Ref.~\cite{PhysRevA.21.2115} but named by the present authors in Ref.~\cite{PhysRevE.101.052124}. It is noted that for the forward CL cycle, $\tilde{t}_p>0$ due to $\Delta S>0$, whereas for the inverse CL cycle, $\tilde{t}_p<0$ due to $\Delta S<0$.

With the help of the new variables~\eqref{append-reduced-matrix}, Eqs.~\eqref{model-leaky-qh}-\eqref{entropy-production} can be rewritten as
\begin{equation}
\frac{\varPhi_{ h}}{T_{ h}\Delta S}=1-\frac{1}{\alpha \tilde{t}_{ p}}+\tilde{\kappa}\tilde{t
}_p(1-\tau)\equiv\tilde{\varPhi}_h, 
\label{ad-1-he-qh}
\end{equation}
\begin{equation}
\frac{\varPhi_{ c}}{T_{ h}\Delta S}=
\tau+\frac{\sigma\tau}{(1-\alpha)\tilde{t}_{ p}}+\tilde{\kappa}\tilde{t
}_p(1-\tau)\equiv\tilde{\varPhi}_c. 
\label{ad-1-he-qc}
\end{equation}
\begin{equation}
    \frac{\Delta S_{tot}}{\Delta S}=\frac{1}{\alpha\tilde{t}_p}+\frac{\sigma}{(1-\alpha)\tilde{t}_p}+\frac{
    \tilde{\kappa}\tilde{t
}_p(1-\tau)^2}{\tau}\equiv\tilde{\Delta S_{tot}}.
    \label{ad-fir-entropy-ad}
\end{equation}

Using further Eqs.~\eqref{ad-1-he-qh}-\eqref{ad-fir-entropy-ad}, together with Eqs.~\eqref{eq:HE-power}-\eqref{eq:HP-eff}, the reduced powers $\{\tilde{P},\tilde{R},\tilde{M}\}$ and the efficiencies $\{\eta,\varepsilon,\epsilon\}$ for LD HEs, REs, and HPs with heat leak can be defined as
\begin{empheq}[left=\text{HE}\empheqlbrace]{align}
\tilde{P}&=\frac{\sigma_h}{T_h\Delta S^2}P=\frac{\tilde{\varPhi}_h-\tilde{\varPhi}_c}{\tilde{t}_{ p}}, \label{eq:HE-power-dimen-fir} \\  
 \eta&=1-\frac{\tilde{\varPhi}_c}{\tilde{\varPhi}_h}=\frac{\eta_C}{1+\frac{\tau\tilde{\Delta S_{tot}}}{\tilde{P}\tilde{t}_p}},\quad\quad \label{eq:HE-eff-dimen-fir} 
\end{empheq} 
\begin{empheq}[left=\text{RE}\empheqlbrace]{align} 
 \tilde{R}&=\frac{\sigma_h}{T_h\Delta S^2}R=\frac{-\tilde{\varPhi}_{ c}}{\tilde{t}_{ p}}, \label{eq:RE-power-dimen-fir} \\  
\varepsilon&=\frac{\tilde{\varPhi}_{ c}}{\tilde{\varPhi}_h-\tilde{\varPhi}_c}=\frac{\varepsilon_C}{1+\frac{\varepsilon_C\Delta \tilde{S}_{tot}}{\tilde{R}\tilde{t}_p}}, \label{eq:RE-eff-dimen-fir}  
\end{empheq}
\begin{empheq}[left=\text{HP}\empheqlbrace]{align} 
 \tilde{M}&=\frac{\sigma_h}{T_h\Delta S^2}M=\frac{-\tilde{\varPhi}_{ h}}{\tilde{t}_{ p}}, \label{eq:HP-power-dimen-fir} \\  
 \epsilon&=\frac{\tilde{\varPhi}_{ h}}{\tilde{\varPhi}_h-\tilde{\varPhi}_c}=\frac{\epsilon_C}{1+\frac{\epsilon_C\tau\Delta \tilde{S}_{tot}}{\tilde{M}\tilde{t}_p}}. \label{eq:HP-eff-dimen-fir}  
\end{empheq}
Each of these quantities depends on the five parameters ${\alpha, \tilde{t}_p, \tilde{\kappa}, \sigma, \tau}$. For future reference, we denote the pairs of power and efficiency for each device by ${\mathcal{A}, \mathcal{B}}$:
\begin{equation}
    \{\mathcal{A},\mathcal{B}\}=
    \begin{cases}
        \{\tilde{P}~\eqref{eq:HE-power-dimen-fir},\eta~\eqref{eq:HE-eff-dimen-fir}\},&\text{HE},\\
        \{\tilde{R}~\eqref{eq:RE-power-dimen-fir},\varepsilon~\eqref{eq:RE-eff-dimen-fir}\},&\text{RE},\\
        \{\tilde{M}~\eqref{eq:HP-power-dimen-fir},\epsilon~\eqref{eq:HP-eff-dimen-fir}\},&\text{HP}.
    \end{cases}
    \label{eq:set-pow-eff}
\end{equation}
Below, we optimize $\mathcal{A}$ and $\mathcal{B}$ under various conditions with respect to the easily adjustable contact times with the individual reservoirs, characterized by $\alpha$ and $\tilde{t}_p$, while assuming that the parameters describing the working medium and its coupling to the heat baths, $\tilde{\kappa}$, $\sigma$, and $\tau$, remain fixed.

\section{Efficiency at maximum power}
\label{sec:emp}

In this section, we investigate the EMP ($\mathcal{B}=\eta,\varepsilon,\epsilon$ at maximum $\mathcal{A}=\tilde{P},\tilde{R},\tilde{M}$) for LD TDs with heat leak. For HEs with heat leak, the EMP has been previously derived in Ref.~\cite{50influence} by direct maximizing power, and in Ref.~\cite{huang2015performance} using the Lagrange Multiplier Method. In Sec.~\ref{sec:EMP-A-HE}, we review the results of Ref.~\cite{50influence}. In Secs.~\ref{sec:EMP-A-RE} and~\ref{sec:EMP-A-HP}, we derive the EMP for REs and HPs with heat leak, which have not been reported so far.

\begin{figure}[t]
\centering
\includegraphics[trim={0cm 0.4cm 0cm 0cm}, width=0.99\columnwidth]{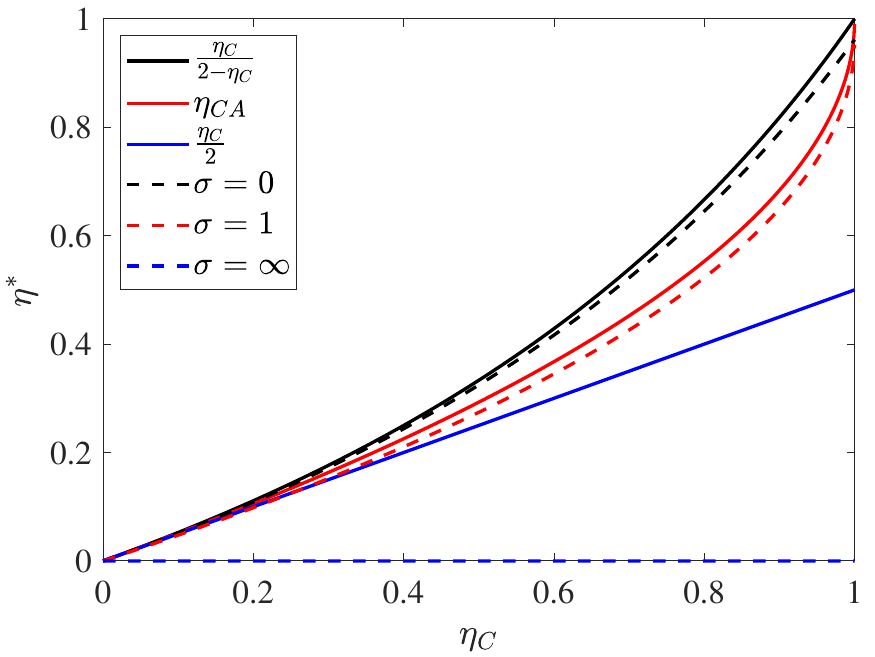}
\caption{Comparison of the EMP, $\eta^*$~\eqref{EMP-HE-fir}, for LD HEs without ($\tilde{\kappa}=0$, solid) and with ($\tilde{\kappa}=0.01$, dashed) heat leak as a function of $\eta_C$. Lines sharing the same color represent the same value of $\sigma$: $\sigma=0$ (black), $\sigma=1$ (red), and $\sigma=\infty$ (blue). Note that the blue dashed line and the horizontal axis overlap.}
\label{fig:EMP-LDHE-leak}
\end{figure}

\subsection{HE}
\label{sec:EMP-A-HE}

Maximizing $\tilde{P}$~\eqref{eq:HE-power-dimen-fir} with respect to $\alpha$ and $\tilde{t}_p$ gives~\cite{PhysRevLett.105.150603,PhysRevE.98.042112,PhysRevE.103.052125,holubec2016maximum,PhysRevE.92.052125}
\begin{equation}
\alpha^*=\frac{1}{1+\sqrt{\sigma\tau}}, 
\label{eq:LDHE-opt-maxP-alpha}
\end{equation}
\begin{equation}
\tilde{t}_p^*=\frac{2}{\eta_C}(1+\sqrt{\sigma\tau})^2,   
\end{equation}
\begin{equation}
\tilde{P}^*=\frac{\eta_C^2}{4(1+\sqrt{\sigma\tau})^2},    
\label{HE-maxP-fir}
\end{equation}
\begin{equation}
\begin{split}
\eta{^*}&=\frac{\eta_C}{2-\frac{\eta_C}{1+\sqrt{\sigma\tau}}+4\tilde{\kappa}(1+\sqrt{\sigma\tau})^2}  \\
&=\frac{\eta_{\rm C}}{2+4\tilde{\kappa}(1+\sqrt{\sigma})^2}+\mathcal{O}(\eta^2_{\rm C}) 
\end{split}.
\label{EMP-HE-fir}
\end{equation}
Due to the independence of $\tilde{P}$~\eqref{eq:HE-power-dimen-fir} from $Q_l$~\eqref{example-fir-exp-ql},
Eqs.~\eqref{eq:LDHE-opt-maxP-alpha}-\eqref{HE-maxP-fir} align with those derived for LD HEs without heat leak~\cite{PhysRevLett.105.150603,PhysRevE.98.042112,PhysRevE.103.052125,holubec2016maximum,PhysRevE.92.052125,huang2015performance}.
However, $Q_l$ decreases the EMP $\eta^*$~\eqref{EMP-HE-fir} by increasing the reduced thermal conductivity $\tilde{\kappa}$. Notably, for $\tilde{\kappa}>0$, the leading term of Taylor expansion of $\eta^*$~\eqref{EMP-HE-fir} falls below $\eta_{\rm C}/2$. Only when $\tilde{\kappa}=0$ is $\eta_{\rm C}/2$ recovered \cite{PhysRevLett.95.190602,PhysRevLett.102.130602}.

For $\sigma=1$ and $\tilde{\kappa}=0$, $\eta^*$~\eqref{EMP-HE-fir} reduces to the Curzon-Ahlborn (CA) efficiency~\cite{curzon1975efficiency}: $\eta^*=\eta_{CA}=1-\sqrt{1-\eta_C}$. Furthermore, $\eta^*$~\eqref{EMP-HE-fir} provides the lower ($\sigma\to\infty$) and upper ($\sigma\to 0$) bounds on the EMP for LD HEs with heat leak~\cite{50influence,huang2015performance}
\begin{equation}
\eta^*_-\equiv\frac{\eta_C}{2+4\tilde{\kappa}\sigma_{\infty}\tau}\le\eta^*\le\frac{\eta_C}{2-\eta_C+4\tilde{\kappa}}.
\label{eq:LDHE-without-EMP-bounds-fir}
\end{equation}
Here, we account for the fact that the result of the limiting process $\sigma \to \infty$, which yields the lower bound, will in practice depend on the relative magnitudes of $\tilde{\kappa}$ and $\sigma$. Formally, we therefore take the limit $\sigma \to \infty$ while keeping $\tilde{\kappa} \sigma$ fixed at a constant value, denoted as $\tilde{\kappa} \sigma_\infty$. The bounds~\eqref{eq:LDHE-without-EMP-bounds-fir} represent a direct generalization of the bounds on EMP, $\frac{\eta_C}{2} \le \eta^* \le \frac{\eta_C}{2-\eta_C}$, previously obtained for $\tilde{\kappa} = 0$~\cite{PhysRevLett.105.150603}.
The above expressions for the EMP are plotted in Fig.~\ref{fig:EMP-LDHE-leak}.


\subsection{RE}
\label{sec:EMP-A-RE}

$\tilde{R}$~\eqref{eq:RE-power-dimen-fir} exhibits a peak with respect to $\tilde{t}_p$ but varies monotonically with changes in $\alpha$, we thus perform the optimization of $\tilde{R}$ in two steps.

First, maximizing $\tilde{R}$~\eqref{eq:RE-power-dimen-fir} with respect to $\tilde{t}_p$ gives
\begin{equation}
\tilde{t}_{p, \alpha}^*=\frac{2\sigma}{\alpha-1},
\label{RE-par-tpa-fir}
\end{equation}
\begin{equation}
\tilde{R}_\alpha^*= \frac{\tau(1-\alpha)}{4\sigma}-\tilde{\kappa}(1-\tau),  
\label{RE-partial-power-fir}
\end{equation}
\begin{equation}
\varepsilon_\alpha^*=\frac{\varepsilon_C-\frac{4\tilde{\kappa}\sigma}{1-\alpha}}{2+\varepsilon_C+\frac{(1-\alpha)(1+\varepsilon_C)}{\alpha\sigma}}.
\label{RE-par-vara-fir}
\end{equation}
Then, one observes that
the partially optimized reduced power~\eqref{RE-partial-power-fir} is a monotonically decreasing function of $\alpha$, whose maximum is attained for $\alpha=\alpha^*=0$ (hot isotherm is much faster than cold isotherm).
In this limit, Eqs.~\eqref{RE-par-tpa-fir} and~\eqref{RE-partial-power-fir} reduce to
\begin{equation}
\tilde{t}_p^*=-2\sigma, 
\label{RE-par-tp-fir-2}
\end{equation}
\begin{equation}
\tilde{R}^*=\frac{\tau}{4\sigma}-\tilde{\kappa}(1-\tau),  
\label{RE-par-Rstar-fir-2}
\end{equation}
As expected below Eqs.~\eqref{append-reduced-matrix}, $\tilde{t}_{p, \alpha}^*<0$ and $\tilde{t}_p^*<0$ in Eqs.~\eqref{RE-par-tpa-fir} and~\eqref{RE-par-tp-fir-2}. 

\begin{figure}[t]
\centering
\includegraphics[trim={0cm 0.4cm 0cm 0cm}, width=0.99\columnwidth]{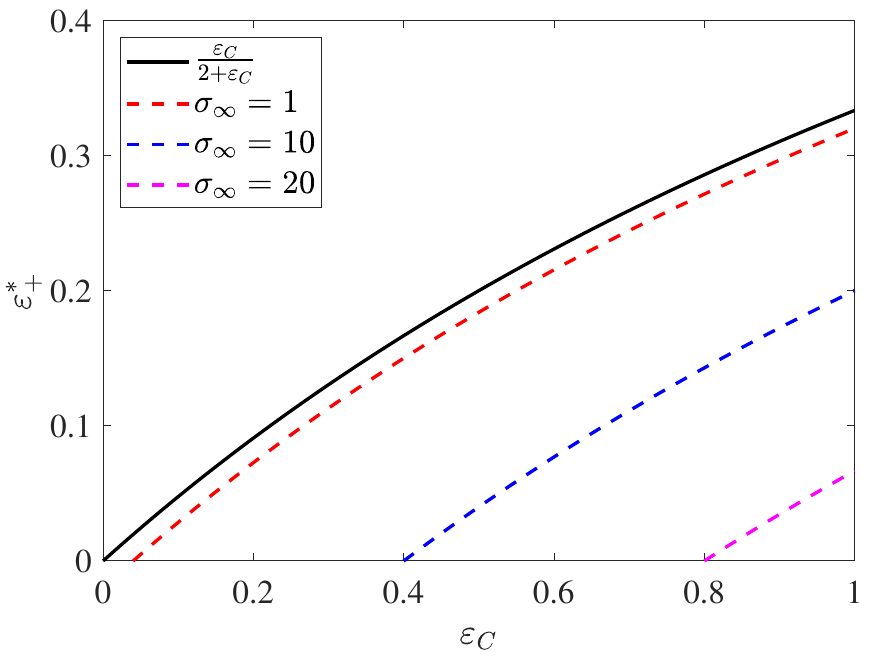}
\caption{Comparison of the upper bound on the EMP, $\varepsilon_+^*$~\eqref{RE-par-etastar-fir-2-bound-1}, for LD REs without ($\tilde{\kappa}=0$, solid) and with ($\tilde{\kappa}=0.01$, dashed) heat leak as a function of $\varepsilon_C$.}
\label{fig:EMP-LDRE-leak}
\end{figure}

Interestingly, when substituting $\alpha=\alpha^*=0$ into Eq.~\eqref{RE-par-vara-fir}, different expressions are obtained, depending on whether $\sigma\to\infty$ (we again fixed $\tilde{\kappa} \sigma$ at a constant value $\tilde{\kappa} \sigma_\infty$ as $\sigma\to\infty$):
\begin{equation}
\varepsilon^*=
\begin{cases}
\frac{\varepsilon_C-4\tilde{\kappa}\sigma_\infty}{2+\varepsilon_C} &\text{for $\sigma\to\infty$}, \\
0
&\text{for finite $\sigma$}. 
\end{cases}   
\label{RE-par-etastar-fir-2}
\end{equation}
This implies that the EMP, $\varepsilon^*$~\eqref{RE-par-etastar-fir-2}, for LD REs with heat leak, experiences a discontinuity at $\sigma\to\infty$ (cold isotherm is `infinitely more irreversible' than the hot one), aligning with the findings in LD REs without heat leak ($\tilde{\kappa}=0$)~\cite{hernandez2015time,PhysRevE.101.052124,PhysRevE.103.052125}.
We conclude from~\eqref{RE-par-etastar-fir-2} that $\varepsilon^*$ is bounded by
\begin{equation}
0\le\varepsilon^*\le \frac{\varepsilon_C-4\tilde{\kappa}\sigma_\infty}{2+\varepsilon_C}\equiv\varepsilon_+^*.
 \label{RE-par-etastar-fir-2-bound-1}
\end{equation}
The upper bound is valid for $\tilde{\kappa}\sigma_\infty < \varepsilon_C/4$, i.e., for small enough heat leak $\tilde{\kappa}$. When $\tilde{\kappa}\sigma_\infty > \varepsilon_C/4$, the upper bound on the efficiency becomes zero.
The inequalities~\eqref{RE-par-etastar-fir-2-bound-1} represent a direct generalization of the bounds on EMP, $0 \leq \varepsilon \leq \frac{\varepsilon_C}{2+\varepsilon_C}$, obtained for $\tilde{\kappa} = 0$~\cite{hernandez2015time,PhysRevE.101.052124,PhysRevE.103.052125}.
The bounds are plotted in Fig.~\ref{fig:EMP-LDRE-leak}


\subsection{HP}
\label{sec:EMP-A-HP}

Like REs, we also have $\tilde{t}_p<0$ for HPs, as predicted below Eqs.~\eqref{append-reduced-matrix}.
Since $\tilde{M}$~\eqref{eq:HP-power-dimen-fir} is a monotonically decreasing function of $\alpha$ and monotonically increasing function of $\tilde{t}_p<0$, it is crucial to first determine their permissible ranges. We rely on the following two conditions: (i) The requirement for positive reduced power, $\tilde{M}>0$~\eqref{eq:HP-power-dimen-fir}, necessitates that $-\tilde{\varPhi}_h<0$; (ii) For the HP to transfer net heat from the cold bath to the hot bath, it is essential that $|-\tilde{\varPhi}_h|>|-\tilde{\varPhi}
_c|$.
These lead to the allowable intervals for $\tilde{t}_p<0$ at fixed $\alpha$,
\begin{widetext}
\begin{equation}
    \frac{1-\epsilon_C}{2\tilde{\kappa}}\left[1+\sqrt{1-\frac{4\tilde{\kappa}\sigma}{(1-\alpha)(\epsilon_C-1)}}\right]<\tilde{t}_p<\frac{1-\epsilon_C}{2\tilde{\kappa}}\left[1-\sqrt{1-\frac{4\tilde{\kappa}\sigma}{(1-\alpha)(\epsilon_C-1)}}\right].
    \label{LD-HP-flash-tp-bound-maxp}
\end{equation}
\end{widetext}

The maximum $\tilde{M}$~\eqref{eq:HP-power-dimen-fir} is attained at minimum and maximum allowed values of $\alpha$ and $\tilde{t}_p<0$,
\begin{equation}
\alpha^*=0,    
\end{equation}
\begin{equation}
\tilde{t}_p^*=\frac{1-\epsilon_C}{2\tilde{\kappa}}\left(1-\sqrt{1-\frac{4\tilde{\kappa}\sigma}{\epsilon_C-1}}\right),  
\label{LD-HP-flash-maxp}
\end{equation}
where $\tilde{t}_p^*$~\eqref{LD-HP-flash-maxp} is obtained by substituting $\alpha=\alpha^*=0$ into the upper bound of $\tilde{t}_p$~\eqref{LD-HP-flash-tp-bound-maxp}.
The resulting maximum reduced power diverges, and the corresponding efficiency attains its smallest possible value:
\begin{equation}
\tilde{M}^*=\infty,    
\end{equation}
\begin{equation} 
\epsilon^*=1. 
\end{equation}
This performance, observed when $\alpha=\alpha^*=0$ (hot isotherm is much faster than cold isotherm), corresponds to $|-\tilde{\varPhi}_h|=|(-W)+(-\tilde{\varPhi}_c)|\gg|-\tilde{\varPhi}_c|$, signifying that the net heat $-\tilde{\varPhi}_c$ pumped from the cold bath is negligible compared to the input work $-W$. Thus, the HP operates as a work-to-heat converter, which is highly undesirable. Note that such performance is also exhibited by LD HPs in the absence of heat leak~\cite{PhysRevE.105.024139}.

\section{Power at maximum efficiency}
\label{sec:PME-fir}

Due to the presence of heat leak, the maximum values of $\eta$, $\varepsilon$, and $\epsilon$ no longer coincide with their respective Carnot limits $\eta_C$, $\varepsilon_C$, and $\epsilon_C$. Therefore, in this section, we
explore the problem of PME ($\mathcal{A}=\tilde{P},\tilde{R},\tilde{M}$ at maximum $\mathcal{B}=\eta,\varepsilon,\epsilon$). 

It is evident from Eqs.~\eqref{eq:HE-eff-dimen-fir},~\eqref{eq:RE-eff-dimen-fir}, and~\eqref{eq:HP-eff-dimen-fir} that 
$\eta$, $\varepsilon$, and $\epsilon$ all solely depend on the ratio $\tilde{\varPhi}_c/\tilde{\varPhi}_h$: (i) Maximum $\eta$ corresponds to minimum $\tilde{\varPhi}_c/\tilde{\varPhi}_h$; (ii) Both maximum $\varepsilon$ and $\epsilon$ correspond to maximum $\tilde{\varPhi}_c/\tilde{\varPhi}_h$. Given that $\eta$, $\varepsilon$, and $\epsilon$ exhibit peak values with respect to $\alpha$ and $\tilde{t}_p$, maximum $\eta$, $\varepsilon$, and $\epsilon$ are determined by the same system of equations,
\begin{equation}
    \begin{cases}
\frac{\partial(\tilde{\varPhi}_c/\tilde{\varPhi}_h)}{\partial{\alpha}}=0,\\
\frac{\partial(\tilde{\varPhi}_c/\tilde{\varPhi}_h)}{\partial{\tilde{t}_p}}=0,
    \end{cases}
\label{PEM-fir-con}
\end{equation}
which lead to
\begin{equation}
\frac{\partial_{\alpha}\tilde{\varPhi}_c}{\partial_{\alpha}\tilde{\varPhi}_h}=\frac{\tilde{\varPhi}_c}{\tilde{\varPhi}_h}=\frac{\partial_{\tilde{t}_p}\tilde{\varPhi}_c}{\partial_{\tilde{t}_p}\tilde{\varPhi}_h}.
\label{PEM-fir-con-namely}
\end{equation}

\begin{figure}[t]
\centering
\includegraphics[trim={0cm 1.5cm 1.6cm 0cm}, width=0.99\columnwidth]{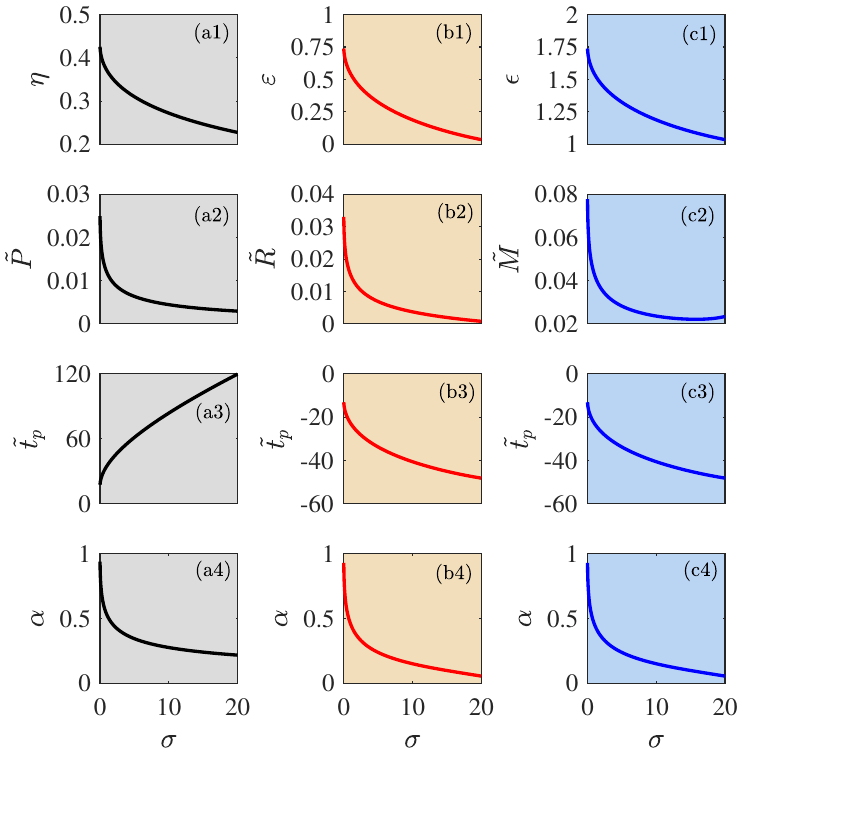}
\caption{Characteristics of LD HEs (first column), REs (second column), and HPs (third column) with heat leak at maximum efficiency as functions of $\sigma$, obtained by numerical solving Eqs.~\eqref{eq:LDHE-PME-FIR-final} and~\eqref{eq:LDHE-PME-SEC-final}. First row: the maximum efficiencies $\mathcal{B}=\eta,\varepsilon,\epsilon$. Second to fourth rows: the corresponding reduced powers $\mathcal{A}=\tilde{P},\tilde{R},\tilde{M}$, and the parameters $\tilde{t}_p$ and $\alpha$. Four observations: (i) As expected, Panels (a1)-(c1) show that $\eta<\eta_C=0.5$, $\varepsilon<\varepsilon_C=1$, and $\epsilon<\epsilon_C=2$; (ii) Consistent with the predictions below Eqs.~\eqref{append-reduced-matrix}, Panel (a3) display positive $\tilde{t}_p$, whereas negative $\tilde{t}_p$ are observed in (b3) and (c3); (iii) In accordance with the analysis above Sec.~\ref{sec:optmal-relation-fir}, $\tilde{t}_p$ and $\alpha$ in Panels (b3) and (b4) aligns with those in (c3) and (c4), respectively; (iv) Panels (b1) and (c1) demonstrate that $\epsilon=\varepsilon+1$. Other parameters used: $\tilde{\kappa}=0.01$ and $\tau=0.5$ (thus $\eta_C=0.5$, $\varepsilon_C=1$, and $\epsilon_C=2$).}
\label{fig:PME-leak}
\end{figure}

Substituting $\tilde{\varPhi}_h$ and $\tilde{\varPhi}_c$ from Eqs.~\eqref{ad-1-he-qh} and~\eqref{ad-1-he-qc} into Eq.~\eqref{PEM-fir-con-namely}, we find the more explicit system of equations determining the maximum $\eta$, $\varepsilon$, and $\epsilon$ for LD HEs, REs, and HPs with heat leak,
\begin{subequations}
\begin{equation}
    \frac{\tilde{\kappa}}{\varepsilon_C}\tilde{t}_p^2+\frac{(1-\alpha)^2-\sigma\alpha^2}{(1-\alpha)^2-\sigma\tau\alpha^2}\tilde{t}_p+\frac{\sigma}{(1-\alpha)^2-\sigma\tau\alpha^2}=0,
    \label{eq:LDHE-PME-FIR-final}
\end{equation}
\begin{equation}
    \eta_C\tilde{t}_p^2-2\left(\frac{1}{\alpha}+\frac{\sigma\tau}{1-\alpha}\right)\tilde{t}_p-\frac{\varepsilon_C}{\tilde{\kappa}}\left(\frac{1}{\alpha}+\frac{\sigma}{1-\alpha}\right)=0.
    \label{eq:LDHE-PME-SEC-final}
\end{equation}
\end{subequations}
Unfortunately, Eqs.~\eqref{eq:LDHE-PME-FIR-final} and~\eqref{eq:LDHE-PME-SEC-final} cannot be solved analytically for $\alpha$ and $\tilde{t}_p$, requiring numerical methods. For HEs, REs, and HPs, the physically reasonable solution for $\alpha$ must lie within the range $0 \leq \alpha \leq 1$. The solution for $\tilde{t}_p$ must be positive for HEs and negative for REs and HPs. Additionally, $\alpha$ and $\tilde{t}_p$ must be identical for REs and HPs, as a consequence of the relation $\epsilon = \varepsilon + 1$, derived from Eqs.~\eqref{eq:RE-eff-dimen-fir} and~\eqref{eq:HP-eff-dimen-fir}.
Numerically obtained solutions to the above equations for HEs, REs, and HPs, satisfying these conditions, are plotted in Fig.~\ref{fig:PME-leak}.

\section{Optimal relation between power and efficiency}
\label{sec:optmal-relation-fir}

In practical applications, the primary interest often lies in delivering a specified amount of power with maximum efficiency, or in maximizing the generated power for given cost, and consequently, the efficiency~\cite{holubec2016maximum,PhysRevE.92.052125}. In this section, we derive the optimal relations between power and efficiency for LD HEs, REs, and HPs with leak. 

\begin{figure*}[t]
\centering
\includegraphics[trim={0cm 3.6cm 8.2cm 0cm}, width=0.99\textwidth]{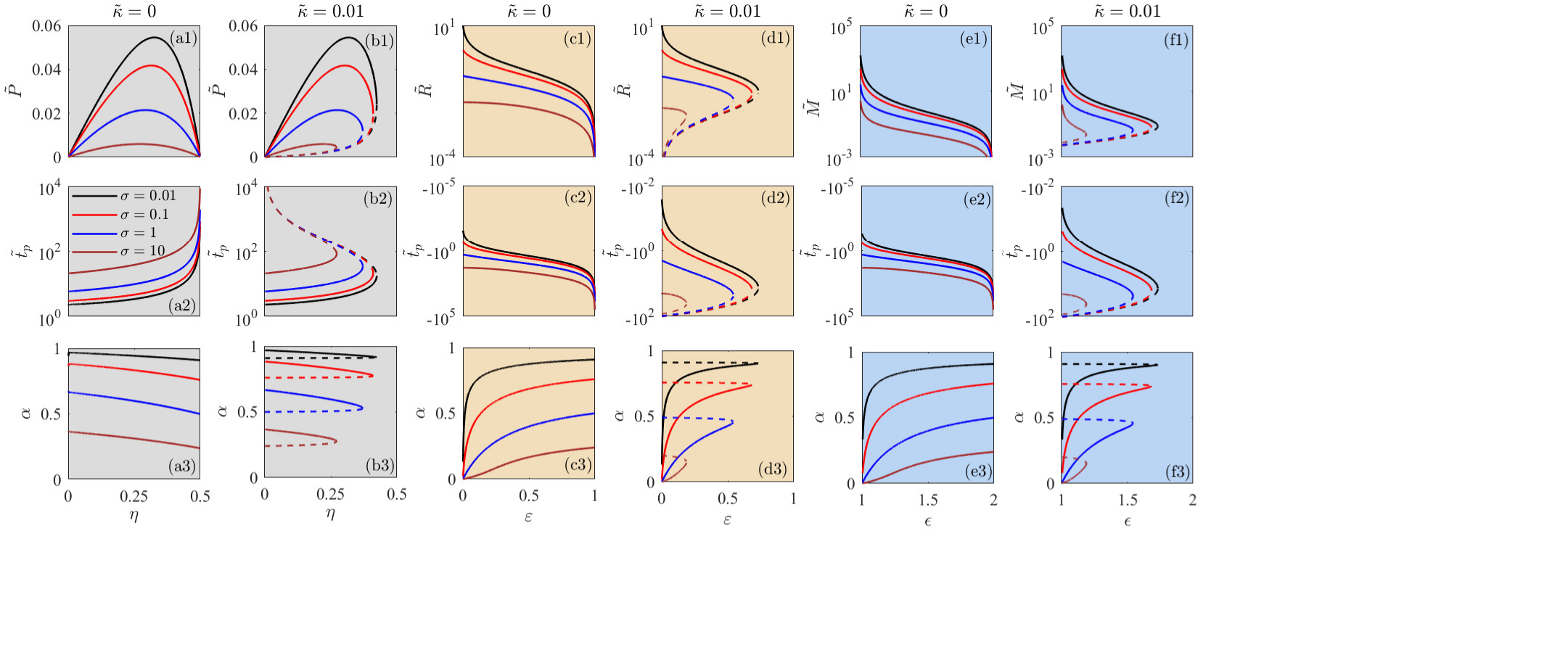}
\caption{Characteristics of LD HEs (left two columns), REs (middle two columns), and HPs (right two columns) without ($\tilde{\kappa}=0$) and with ($\tilde{\kappa}=0.01$) heat leak as functions of the efficiencies $\mathcal{B}=\eta,\varepsilon,\epsilon$. First row: the upper (solid) and lower (dashed) bounds of the reduced powers, $\mathcal{A}=\tilde{P},\tilde{R},\tilde{M}$. Second and third rows: the corresponding parameters $\tilde{t}_p$ and $\alpha$. For $\tilde{\kappa}=0$, the lower bounds of $\mathcal{A}=\tilde{P},\tilde{R},\tilde{M}$ are 0. Thus, the dashed lines are not plotted in the first, third, and fifth columns. Other parameters used: $\tau=0.5$ and $\sigma=0.01,0.1,1,10$.}
\label{fig:MPGE-fir}
\end{figure*}

\begin{figure*}[t]
\centering
\includegraphics[trim={0cm 3.9cm 8.5cm 0cm}, width=0.99\textwidth]{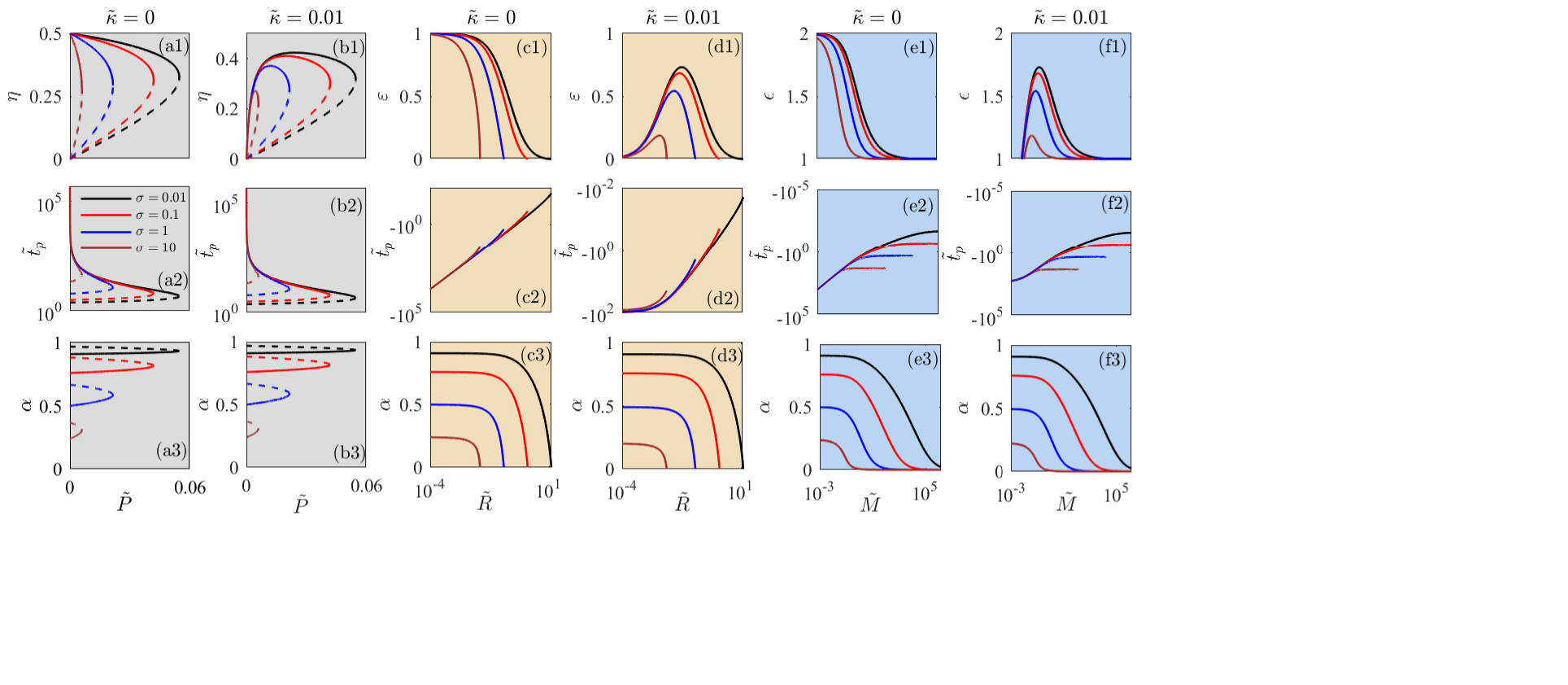}
\caption{Characteristics of LD HEs (left two columns), REs (middle two columns), and HPs (right two columns) without ($\tilde{\kappa}=0$) and with ($\tilde{\kappa}=0.01$) heat leak as functions of the reduced powers $\mathcal{A}=\tilde{P},\tilde{R},\tilde{M}$. First row: the upper (solid) and lower (dashed) bounds of efficiency, $\mathcal{B}=\eta,\varepsilon,\epsilon$. Second and third rows: the corresponding parameters $\tilde{t}_p$ and $\alpha$. The lower bounds of $\varepsilon$ and $\epsilon$ are given by their smallest possible values, 0 and 1. Thus, the dashed lines are not plotted in the right four columns. Other parameters used: $\tau=0.5$ and $\sigma=0.01,0.1,1,10$.}
\label{fig:MEGP-fir}
\end{figure*}

Available results on BPFE and BEFP~\cite{zhao2022microscopic,yan1990class,Chen1989TheEO, chen1995study} suggest that, despite being constrained optimization problems with different constraints, optimizing power at fixed efficiency and optimizing efficiency at fixed power ultimately yield identical results. In the present setup, this can be verified by plotting the numerically obtained bounds and comparing the results, as shown in the first rows of Figs.~\ref{fig:MPGE-fir} and~\ref{fig:MEGP-fir}. Analytically, the conjecture has been proven for LD HEs constructed via shortcut strategies by Zhao \emph{et al}.~\cite{zhao2022microscopic}. We now provide a more general proof based on the method of Lagrange multipliers~\cite{chen2021microscopic,Chen1989TheEO,dechant2017underdamped,huang2015performance,ye2017universality}, which shows that the conjecture is valid for arbitrary model.

Consider two functions, $\mathcal{A} = \mathcal{A}(p_1, \dots, p_n)$ and $\mathcal{B} = \mathcal{B}(p_1, \dots, p_n)$, which depend on parameters $p_i$ with $i = 1, \dots, n$. Assuming that $\mathcal{A}$ and $\mathcal{B}$ are smooth functions of their parameters, the optimization of $\mathcal{A}$ for fixed $\mathcal{B}$ can be performed by finding the stationary points $\partial_{p_i} L_A = 0$ of the Lagrangian $L_A = \mathcal{A} + \lambda \mathcal{B}$. Taking the partial derivatives yields $\partial_{p_i} \mathcal{A} + \lambda  \partial_{p_i} \mathcal{B} = 0$ for all $i$. Solving these $n$ equations for the Lagrange multiplier $\lambda$ and equating the results leads to $n-1$ equations of the form 
\begin{equation} \frac{\partial_{p_i} \mathcal{A}}{\partial_{p_i} \mathcal{B}} = \frac{\partial_{p_{i+1}} \mathcal{A}}{\partial_{p_{i+1}} \mathcal{B}}, \label{eq:set} \end{equation} 
with $i = 1, \dots, n-1$. We assumed that the denominators are nonzero. Importantly, fixing $\mathcal{A}$ and optimizing $\mathcal{B}$ results in the same set of equations through an analogous calculation.

The constraint $\mathcal{B}(p_1, \dots, p_n) = \mathcal{B}$ can be used to express one of the parameters, for example $p_1 = p_1(p_2, \dots, p_n, \mathcal{B})$, while the remaining parameters are determined by solving Eqs.~\eqref{eq:set}, yielding their optimal values and thus the extremal (maximum or minimum) value of $\mathcal{A}$ for the fixed $\mathcal{B}$.
Fixing $\mathcal{A}$ and optimizing $\mathcal{B}$ leads to the same set of optimal parameters, provided that the solution for $p_1$ from $\mathcal{A}(p_1, \dots, p_n) = \mathcal{A}$ coincides with that obtained when optimizing $\mathcal{B}$.
By varying $\mathcal{B}$ over its admissible range while optimizing $\mathcal{A}$, one sweeps through all physically allowed values of $p_1$, and similarly by varying $\mathcal{A}$ while optimizing $\mathcal{B}$. Hence, plotting the bounds on $\mathcal{A}$ at fixed $\mathcal{B}$ yields the same curve in the $\mathcal{A}$–$\mathcal{B}$ diagram as plotting the bounds on $\mathcal{B}$ at fixed $\mathcal{A}$.

Now, let us substitute Eqs.~\eqref{eq:HE-power-dimen-fir}–\eqref{eq:HP-eff-dimen-fir} into Eq.~\eqref{eq:set}, where $p_1 = \alpha$, $p_2 = \tilde{t}_p$, and $n = 2$.
The optimal relationships between $\alpha$ and $\tilde{t}_p$ for the considered LD HEs, REs, and HPs with heat leak, derived by solving Eq.~\eqref{eq:set}, are identical. This results from the fact that these devices—HEs, REs, and HPs—represent different operational modes of a single system, and their optimization can be mapped to the optimization of the average entropy production rate, as demonstrated in Sec.\ref{sec:relation-bounds-fir}. The optimal relationship reads
\begin{equation}
    \tilde{t}_p=\frac{2 \sigma }{\alpha ^2 (\sigma -1)+2 \alpha -1}.
\label{append-fir-relation-sec-LD}
\end{equation}
Surprisingly, it is independent of $\tilde{\kappa}$. This implies that Eq.~\eqref{append-fir-relation-sec-LD} is applicable regardless of whether the heat leak is taken into account. Given the conditions that $\tilde{t}_p>0$ for the forward CL cycle and $\tilde{t}_p<0$ for the inverse CL cycle, combined with the constraint $\alpha\in[0,1]$, we can further deduce from Eq.~\eqref{append-fir-relation-sec-LD} the following permissible ranges for $\alpha$ and $\tilde{t}_p$:
\begin{align}
    \alpha\in\left[\frac{1}{1+\sqrt{\sigma}}, 1\right]~\&~\tilde{t}_p\in[2,\infty],&~\text{HE},
    \label{permis-1}
    \\
    \alpha\in\left[0,\frac{1}{1+\sqrt{\sigma}}\right]~\&~\tilde{t}_p\in[-\infty,-2\sigma],&~\text{RE \& HP}.
    \label{permis-2}
\end{align}

Inserting Eq.~\eqref{append-fir-relation-sec-LD} into $\mathcal{A} = \mathcal{A}(\alpha, \tilde{t}_p)$ and $\mathcal{B} = \mathcal{B}(\alpha, \tilde{t}_p)$, we obtain the parametric equations
\begin{align}
\mathcal{A}&=\mathcal{A}\left[\alpha,\frac{2 \sigma }{\alpha ^2 (\sigma -1)+2 \alpha -1}\right],
    \label{append-fir-relation-opt-a-tp-fir-fir}\\
\mathcal{B}&=\mathcal{B}\left[\alpha,\frac{2 \sigma }{\alpha ^2 (\sigma -1)+2 \alpha -1}\right],
    \label{append-fir-relation-opt-a-tp-sec-sec}
\end{align}
which represent the optimal relation between power and efficiency for LD HEs, REs, and HPs with heat leak. 
To explicitly derive the BPFE from Eqs.~\eqref{append-fir-relation-opt-a-tp-fir-fir} and~\eqref{append-fir-relation-opt-a-tp-sec-sec}, we need to express $\mathcal{A}$ in terms of $\mathcal{B}$. Conversely, to explicitly obtain the BEFP, we have to express $\mathcal{B}$ in terms of $\mathcal{A}$. Assuming that the parameters $\tilde{\kappa}$, $\sigma$, and $\tau$ incorporated in Eqs.~\eqref{append-fir-relation-opt-a-tp-fir-fir} and~\eqref{append-fir-relation-opt-a-tp-sec-sec} are given, for both cases, we only need to eliminate the remaining common parameter $\alpha$. However, due to the complex dependencies of $\mathcal{A}$ and $\mathcal{B}$ in Eqs.~\eqref{append-fir-relation-opt-a-tp-fir-fir} and~\eqref{append-fir-relation-opt-a-tp-sec-sec} on $\alpha$, it is generally not possible to express $\mathcal{A}$ in terms of $\mathcal{B}$ and/or $\mathcal{B}$ in terms of $\mathcal{A}$. To this end, in Sec.~\ref{sec:Explicit}, we delve into the special parameter regimes where explicit expressions for BPFE and BEFP can be obtained. In Sec.~\ref{sec:Numerical}, we numerically investigate these bounds for arbitrary parameters.

\subsection{Explicit expressions}
\label{sec:Explicit}

\textbf{(i)} $\bm{\tilde{\kappa}=0}.$ In this limit, the BPFE and BEFP for LD HEs~\cite{holubec2016maximum,PhysRevLett.124.110606,PhysRevE.103.052125,PhysRevE.98.042112,zhao2022microscopic,PhysRevE.106.054108}, REs~\cite{abiuso2020geometric,PhysRevE.101.052124,PhysRevE.103.052125}, and HPs~\cite{PhysRevE.105.024139} have been extensively studied. In particular, Ma and Fu~\cite{ma2024unified} recently proposed a unified framework to analyze the power-efficiency trade-off in generic TDs. While explicit expressions for the BPFE and/or BEFP are attainable for $\tilde{\kappa}=0$~\cite{PhysRevLett.124.110606,PhysRevE.98.042112,PhysRevE.101.052124,PhysRevE.103.052125,PhysRevE.105.024139}, they are too lengthy and complex to present here.
    
\textbf{(ii)} $\bm{\sigma=1.}$
It is evident from Eq.~\eqref{append-fir-relation-sec-LD} that, in the symmetric dissipation ($\sigma=1$), Eq.~\eqref{append-fir-relation-sec-LD} simplifies, and consequently, so do Eqs.~\eqref{append-fir-relation-opt-a-tp-fir-fir} and~\eqref{append-fir-relation-opt-a-tp-sec-sec}. In this limit, the lower (upper sign) and upper (lower sign) BEFP for LD HEs with heat leak can be expressed as
\begin{equation}
\small{
    \eta=\frac{8\tilde{P}}{4\tilde{P}+\eta_C(1+8\tilde{\kappa})\pm\sqrt{8\tilde{P}(2\tilde{P}+\eta_C)+\eta_C^2-16\tilde{P}}}}.
    \label{simp-LD-HE-1}
\end{equation}
For REs and HPs with leak, the upper BEFP are, respectively, given by
\begin{equation}   \varepsilon=\frac{(\tilde{R}+\tilde{R}\varepsilon_C)\left[\varepsilon_C-4\tilde{R}(1+\varepsilon_C)-4\tilde{\kappa}\right]}{(\tilde{R}+\tilde{R}\varepsilon_C+\tilde{\kappa})\left[1+4\tilde{R}(1+\varepsilon_C)+4\tilde{\kappa}\right]},
\label{simp-LD-RE-1}
\end{equation}
\begin{equation} \epsilon=\frac{\tilde{M}\epsilon_C(\epsilon_C+4\tilde{M}\epsilon_C+4\tilde{\kappa})}{(\tilde{M}\epsilon_C+\tilde{\kappa})(1+4\tilde{M}\epsilon_C+4\tilde{\kappa})}.
\label{simp-LD-HP-1}
\end{equation}
For REs, the lower BEFP is $\varepsilon=0$, while for HPs, the lower bound is $\epsilon=1$.

\textbf{(iii)} $\bm{\tilde{\kappa}=0~\&~\sigma=1.}$
Eliminating the common parameter $\alpha$ in Eqs.~\eqref{append-fir-relation-opt-a-tp-fir-fir} and~\eqref{append-fir-relation-opt-a-tp-sec-sec} produces the upper BPFE~\cite{PhysRevLett.124.110606,zhao2022microscopic,huang2015performance,PhysRevE.106.054108}
\begin{equation}
    \tilde{P}=\frac{\eta(\eta_C-\eta)}{4(1-\eta)},
    \label{simp-LD-HE-10}
\end{equation}
\begin{equation}
    \tilde{R}=\frac{\varepsilon_C-\varepsilon}{4(1+\varepsilon)(1+\varepsilon_C)},
    \label{simp-LD-RE-10}
\end{equation}
\begin{equation}
    \tilde{M}=\frac{\epsilon_C-\epsilon}{4\epsilon_C(1-\epsilon)},
    \label{simp-LD-HP-10}
\end{equation}
and the lower BPFE $\tilde{P}=\tilde{R}=\tilde{M}=0$. Note that Eqs.~\eqref{simp-LD-HE-10}-\eqref{simp-LD-HP-10} can also be obtained by solving Eqs.~\eqref{simp-LD-HE-1}-\eqref{simp-LD-HP-1}, highlighting the fact that BPFE and BEFP are described by the same expression.

\subsection{Numerical results}
\label{sec:Numerical}

Outside the regimes discussed above, the common parameter $\alpha$ in Eqs.~\eqref{append-fir-relation-opt-a-tp-fir-fir} and~\eqref{append-fir-relation-opt-a-tp-sec-sec} cannot be eliminated analytically. Consequently, we have investigated the BPFE and BEFP, using numerical methods. The results—shown in the first row of Figs.~\ref{fig:MPGE-fir} and~\ref{fig:MEGP-fir}, respectively—lead to three main observations:

(i) The presence of heat leak qualitatively alters the power–efficiency characteristics of HEs, REs, and HPs. For HEs, the power–efficiency curve exhibits an open parabolic shape when $\tilde{\kappa}=0$, and transforms into a closed loop when $\tilde{\kappa}=0.01$, consistent with the findings of Refs.~\cite{huang2015performance,PhysRevE.91.052140}. Such closed-loop curves are typical of real TDs, which inevitably experience external sources of irreversibility, such as heat leaks~\cite{chen1994maximum,chen2001curzon,PhysRevE.91.050102}.
For REs and HPs, the power–efficiency curves are monotonic in the absence of heat leak ($\tilde{\kappa}=0$), but become parabolic when heat leak is present ($\tilde{\kappa}=0.01$). 

Intuitively, this result can be understood as follows: in the absence of heat leak, the minimum power at given efficiency is necessarily zero for all devices. This corresponds to vanishing heat fluxes, such that the ratio of power to input heat—which defines the efficiency—still yields the desired fixed value. However, when a heat leak is present, zero power can no longer produce a finite efficiency, since the heat leak makes the input heat necessarily nonzero. As a result, the minimum power at fixed efficiency becomes finite, leading to two possible power values for each efficiency in the plot.

(ii) Whenever the upper BPFE is concave—such that the slope of the bound at maximum power vanishes—a slight reduction of power from its maximum value leads to a significant increase in efficiency~\cite{PhysRevLett.112.130601,PhysRevB.91.115425,PhysRevE.93.050101,holubec2016maximum,dechant2017underdamped,PhysRevE.101.052124,PhysRevE.105.024139}. A similar behavior is observed for the BEFP.
For power, this effect is evident in HEs both with and without heat leak, as shown in Figs.~\ref{fig:MPGE-fir}~(a1) and (b1).
For efficiency, the effect occurs in HEs, REs, and HPs with heat leak [Figs.~\ref{fig:MEGP-fir}~(b1), (d1), and (f1)], but not in the absence of heat leak. Since the reported qualitative shapes of the power–efficiency curves are universal, the observed gains in power and efficiency should be applicable to all TDs, regardless of specific details.

(iii) As shown at the beginning of Sec.\ref{sec:optmal-relation-fir}, the BPFE and BEFP are mathematically equivalent. This equivalence can be visually confirmed by rotating the first row of Fig.\ref{fig:MPGE-fir} clockwise by $90^\circ$ and reflecting the result across the horizontal axis. However, the equivalence of the upper BPFE and BEFP (solid lines in Figs.~\ref{fig:MPGE-fir} and~\ref{fig:MEGP-fir}) hinges on whether PME and EMP are located on the horizontal axis, i.e., coincide with their minimum possible values. Specifically, there are four distinct scenarios:
\begin{itemize}
    \item When both PME and EMP coincide with their minimum possible values, the upper BPFE and BEFP completely overlap [cf.~Panels (c1) and (e1) in Figs.~\ref{fig:MPGE-fir} and~\ref{fig:MEGP-fir}].
    \item When PME exceeds its minimum possible value while EMP remains at its minimum, the upper BPFE is a subset of the upper BEFP [cf.~Panels (d1) and (f1) in Figs.~\ref{fig:MPGE-fir} and~\ref{fig:MEGP-fir}].
    \item When EMP exceeds its minimum possible value while PME remains at its minimum, the upper BEFP is a subset of the upper BPFE [cf.~Panels (a1) in Figs.~\ref{fig:MPGE-fir} and~\ref{fig:MEGP-fir}].
    \item When both PME and EMP exceed their minimum possible values, the upper BPFE and BEFP partly overlap [cf.~Panels (b1) in Figs.~\ref{fig:MPGE-fir} and~\ref{fig:MEGP-fir}].
\end{itemize}
A similar discussion applies to the lower BPFE and BEFP (dashed lines in Figs.~\ref{fig:MPGE-fir} and~\ref{fig:MEGP-fir}).

The control parameters $\tilde{t}_p$ and $\alpha$ corresponding to the BPFE and BEFP are plotted in the second and third rows of Figs.~\ref{fig:MPGE-fir} and~\ref{fig:MEGP-fir}. One can observe that: (i) As expected, $\tilde{t}_p > 0$ for heat engines (HEs), while $\tilde{t}_p < 0$ for refrigerators (REs) and heat pumps (HPs). The parameters $\tilde{t}_p$ and $\alpha$ always satisfy the conditions~\eqref{permis-1} and~\eqref{permis-2}; (ii) The presence of heat leak alters the shapes of $\tilde{t}_p$ and $\alpha$ in Fig.~\ref{fig:MPGE-fir}, whereas it does not appreciably change their shapes in Fig.~\ref{fig:MEGP-fir}; (iii) In Fig.~\ref{fig:MPGE-fir}, $\tilde{t}_p$ and $\alpha$ are the same for REs and HPs. This holds because fixing $\varepsilon$ is equivalent to fixing $\varepsilon = \epsilon + 1$, and, according to Eqs.~\eqref{PRM-array} in the next section, maximizing the power $\tilde{R}$ at fixed efficiency $\varepsilon$ is mathematically equivalent to maximizing power $\tilde{M}$ at fixed efficiency $\epsilon$. By contrast, in Fig.~\ref{fig:MEGP-fir}, $\tilde{t}_p$ and $\alpha$ differ for REs and HPs, since no such relation exists between the powers $\tilde{R}$ and $\tilde{M}$ as exists between the efficiencies.

\section{Optimization Framework for Thermal Machines}
\label{sec:relation-bounds-fir}

Let us now investigate the link among the optimal relations between power and efficiency for LD HEs, REs, and HPs with heat leak derived in Sec.~\ref{sec:optmal-relation-fir}. This link serves to reduce the number of calculations, enabling the derivation of all optimal relations from a single result.

To this end, combining Eqs.~\eqref{eq:HE-eff-dimen-fir},~\eqref{eq:RE-eff-dimen-fir}, and~\eqref{eq:HP-eff-dimen-fir} gives
\begin{equation}
 \left[
 \begin{array}{ccc}
  \tilde{P}   \\
  \tilde{R}   \\
  \tilde{M} 
\end{array}   
\right]=
\frac{\Delta \tilde{S}_{\rm tot}}{\tilde{t}_p}
 \left[
 \begin{array}{ccc}
 \tau\left(\frac{\eta_{\rm C}}{\eta}-1\right)^{-1}   \\
  \left(\frac{1}{\varepsilon}-\frac{1}{\varepsilon_{\rm C}}\right)^{-1}   \\
  \tau\left(\frac{1}{\epsilon}-\frac{1}{\epsilon_{\rm C}}\right)^{-1} 
\end{array}   
\right],
\label{PRM-array}
\end{equation}
where the ratio $\Delta \tilde{S}_{\rm tot}/\tilde{t}_{ p}$ denotes the reduced (average) entropy production rate per cycle. It can be seen from Eqs.~\eqref{PRM-array} that both optimizing power at fixed efficiency ($\mathcal{A}=\tilde{P},\tilde{R},\tilde{M}$ at fixed $\mathcal{B}=\eta,\varepsilon,\epsilon$) and optimizing efficiency at fixed power ($\mathcal{B}=\eta,\varepsilon,\epsilon$ at fixed $\mathcal{A}=\tilde{P},\tilde{R},\tilde{M}$) are equivalent to optimizing $\Delta \tilde{S}_{\rm tot}/\tilde{t}_{ p}$. For the considered LD HEs, REs, and HPs with heat leak, the expressions for $\Delta \tilde{S}_{\rm tot}/\tilde{t}_{ p}$~\eqref{ad-fir-entropy-ad} are the same. Therefore, it is not surprising that the control parameters $\tilde{t}_p$ and $\alpha$ corresponding to the BPFE are consistent for REs and HPs, as shown in Fig.~\ref{fig:MPGE-fir}.
Furthermore, Eqs.~\eqref{PRM-array} show that maximizing $\mathcal{A}$ at fixed $\mathcal{B}$ and minimizing $\mathcal{B}$ at fixed $\mathcal{A}$ are both equivalent to maximizing the reduced average entropy production rate during the cycle, $\Delta \tilde{S}_{\rm tot}/\tilde{t}_{p}$. Importantly, a similar relation can be derived for general CL devices using Eqs.~\eqref{entropy-production}--\eqref{eq:HP-eff}, and as a result, the above conclusions hold for these machines in general.


Eliminating $\Delta \tilde{S}_{\rm tot}/\tilde{t}_{ p}$ in Eqs.~\eqref{PRM-array} yields~\cite{chen1995study,raux2025three}
\begin{equation}
\left(
\begin{array}{cc}
-\tilde{R} & \varepsilon  \\
-\tilde{M} & \epsilon
\end{array}  
\right)=
\left(
\begin{array}{cc}
\frac{\tilde{P}}{\eta}-\tilde{P} & \frac{1}{\eta}-1  \\
\frac{\tilde{P}}{\eta}     & \frac{1}{\eta}
\end{array}
\right),
\label{matrix-1-opt}
\end{equation}
Equations~\eqref{matrix-1-opt} embody the link among the optimal relations we seek, and this connection can be most readily verified by employing Eqs.~\eqref{simp-LD-HE-1}--\eqref{simp-LD-HP-1} and Eqs.~\eqref{simp-LD-HE-10}--\eqref{simp-LD-HP-10}. The relation expressed in Eq.~\eqref{matrix-1-opt} is of practical significance, as it enables engineers to deduce the optimal performance of all three devices---HEs, REs, and HPs---by optimizing only one of them. This optimization, however, must also be performed outside the standard operation regime, where the output power is positive (see Fig.~\ref{fig:model}). Furthermore, Eqs.~\eqref{matrix-1-opt} can be directly derived from the general expressions in Eqs.~\eqref{eq:HE-power}--\eqref{eq:HP-eff}, demonstrating their applicability to general CL TDs with heat leak~\cite{curzon1975efficiency,yan1990class}. The results in Eqs.~\eqref{PRM-array} and \eqref{matrix-1-opt} are a natural consequence of the fact that HEs, REs, and HPs represent different operation modes of a single TD~\cite{raux2025three}.


In closing, we give a remark on the comparison
of LD and endoreversible models, which was done in Refs.~\cite{PhysRevE.96.012151,PhysRevE.102.012151,PhysRevE.105.024139,PhysRevE.97.022139,gonzalez2017carnot,guo2023performance,gonzalez2016irreversible}. In our previous work~\cite{PhysRevE.105.024139}, we demonstrated that, under specific constraints on the control parameters, the upper BEFP for LD and endoreversible TDs are identical in the absence of heat leak. Here, we further tested that the same constraints also result in identical expressions for the lower BEFP, regardless of whether the heat leak is taken into account (data not shown).
The reason why the same constraints are applicable to HEs, REs, and HPs can be traced to the fact that HEs and REs (or HPs) are the same device operating in different regimes.

\section{Conclusion}
\label{sec-con}

In summary, we presented a unified description of CL HEs, REs, and HPs with heat leak. We optimized LD CL HEs, REs, and HPs with heat leak from the perspectives of EMP, PME, BPFE, and BEFP. We provided general proof that the BPFE and BEFP form identical curves in the power-efficiency diagram for arbitrary TDs. Additionally, we derived a general relation that allows for the unified optimization of power and efficiency in CL HEs, REs, and HPs by optimizing the average entropy production rate per cycle. The validity of this relation arises from the fact that HEs and REs (or HPs) are essentially the same device operating in different regimes. Our results for LD TDs with heat leak can be directly applied to calculate the corresponding results for minimally nonlinear irreversible models via the mapping between the two frameworks~\cite{izumida2012efficiency,PhysRevE.91.052140}.

The present study admits several avenues for generalization. A key extension involves establishing connections among the performance of thermal devices (TDs) with respect to alternative figures of merit~\cite{PhysRevE.85.010104, PhysRevE.63.037102, angulo1991ecological, PhysRevE.87.012105, PhysRevE.82.051101, PhysRevE.93.032152, apertet2013efficiency}. The link between the optimal relations of power and efficiency has been verified using endoreversible models without heat leak~\cite{chen1995study} and, in this work, by low-dissipation (LD) models. It is thus expected to also hold for linear irreversible systems~\cite{PhysRevLett.112.180603, PhysRevLett.95.190602, iyyappan2020efficiency}, as well as minimally nonlinear irreversible models~\cite{PhysRevE.91.052140, izumida2012efficiency, izumida2013coefficient} through their established mapping to the LD framework~\cite{izumida2012efficiency, PhysRevE.91.052140, PhysRevLett.95.190602}. Nevertheless, the general validity of this link in more complex settings, as suggested by our proof, remains to be confirmed.
Moreover, leveraging the uncovered connection, the optimization of absorption REs~\cite{PhysRevE.103.052125}---comprising simultaneously operating HEs and REs---and absorption HPs~\cite{guo2020equivalent}---formed by jointly operating HEs and HPs---is expected to be substantially simplified.
Finally, further internal connections among HEs, REs, and HPs remain to be investigated in future work.

\begin{acknowledgments}
ZY acknowledges the supports of Jiangxi Provincial Natural Science Foundation of China (Grant No. 20232BAB211027), the Special Project for Cultivating Early-Career Youth Talents in Science and Technology of Jiangxi Province (Grant No. 20244BCE52030), the Research Startup Foundation for PhD of Jiangxi Normal University (Grant No.~12023513), and China Scholarship Council (CSC) (Grant No.~201906310136). VH acknowledges the funding provided by Charles University through Project PRIMUS/22/SCI/009.
\end{acknowledgments}

\appendix

\bibliography{References}

\begin{thebibliography}{75}%
\makeatletter
\providecommand \@ifxundefined [1]{%
 \@ifx{#1\undefined}
}%
\providecommand \@ifnum [1]{%
 \ifnum #1\expandafter \@firstoftwo
 \else \expandafter \@secondoftwo
 \fi
}%
\providecommand \@ifx [1]{%
 \ifx #1\expandafter \@firstoftwo
 \else \expandafter \@secondoftwo
 \fi
}%
\providecommand \natexlab [1]{#1}%
\providecommand \enquote  [1]{``#1''}%
\providecommand \bibnamefont  [1]{#1}%
\providecommand \bibfnamefont [1]{#1}%
\providecommand \citenamefont [1]{#1}%
\providecommand \href@noop [0]{\@secondoftwo}%
\providecommand \href [0]{\begingroup \@sanitize@url \@href}%
\providecommand \@href[1]{\@@startlink{#1}\@@href}%
\providecommand \@@href[1]{\endgroup#1\@@endlink}%
\providecommand \@sanitize@url [0]{\catcode `\\12\catcode `\$12\catcode `\&12\catcode `\#12\catcode `\^12\catcode `\_12\catcode `\%12\relax}%
\providecommand \@@startlink[1]{}%
\providecommand \@@endlink[0]{}%
\providecommand \url  [0]{\begingroup\@sanitize@url \@url }%
\providecommand \@url [1]{\endgroup\@href {#1}{\urlprefix }}%
\providecommand \urlprefix  [0]{URL }%
\providecommand \Eprint [0]{\href }%
\providecommand \doibase [0]{https://doi.org/}%
\providecommand \selectlanguage [0]{\@gobble}%
\providecommand \bibinfo  [0]{\@secondoftwo}%
\providecommand \bibfield  [0]{\@secondoftwo}%
\providecommand \translation [1]{[#1]}%
\providecommand \BibitemOpen [0]{}%
\providecommand \bibitemStop [0]{}%
\providecommand \bibitemNoStop [0]{.\EOS\space}%
\providecommand \EOS [0]{\spacefactor3000\relax}%
\providecommand \BibitemShut  [1]{\csname bibitem#1\endcsname}%
\let\auto@bib@innerbib\@empty
\bibitem [{\citenamefont {Marks}(2007)}]{marks2007origins}%
  \BibitemOpen
  \bibfield  {author} {\bibinfo {author} {\bibfnamefont {R.}~\bibnamefont {Marks}},\ }\href {https://books.google.cz/books?id=HMdoi0rGmmwC} {\emph {\bibinfo {title} {The Origins of the Modern World: A Global and Ecological Narrative from the Fifteenth to the Twenty-first Century}}},\ G - Reference, Information and Interdisciplinary Subjects Series\ (\bibinfo  {publisher} {Rowman \& Littlefield Publishers},\ \bibinfo {year} {2007})\BibitemShut {NoStop}%
\bibitem [{\citenamefont {Th{\'e}venot}(1979)}]{thévenot1979history}%
  \BibitemOpen
  \bibfield  {author} {\bibinfo {author} {\bibfnamefont {R.}~\bibnamefont {Th{\'e}venot}},\ }\href {https://books.google.cz/books?id=2c1aswEACAAJ} {\emph {\bibinfo {title} {A History of Refrigeration Throughout the World}}}\ (\bibinfo  {publisher} {International Institute of Refrigeration},\ \bibinfo {year} {1979})\BibitemShut {NoStop}%
\bibitem [{\citenamefont {Pearson}(2022)}]{Pearson2022}%
  \BibitemOpen
  \bibfield  {author} {\bibinfo {author} {\bibfnamefont {A.}~\bibnamefont {Pearson}},\ }\bibfield  {title} {\bibinfo {title} {Development of refrigeration and heat pump systems},\ }\href {https://www.frontiersin.org/journals/thermal-engineering/articles/10.3389/fther.2022.1042347} {\bibfield  {journal} {\bibinfo  {journal} {Frontiers in Thermal Engineering}\ }\textbf {\bibinfo {volume} {Volume 2 - 2022}} (\bibinfo {year} {2022})}\BibitemShut {NoStop}%
\bibitem [{Tu_(2012)}]{Tu_2012}%
  \BibitemOpen
  \bibfield  {title} {\bibinfo {title} {Recent advance on the efficiency at maximum power of heat engines},\ }\href {https://doi.org/10.1088/1674-1056/21/2/020513} {\bibfield  {journal} {\bibinfo  {journal} {Chin. Phys. B}\ }\textbf {\bibinfo {volume} {21}},\ \bibinfo {pages} {020513} (\bibinfo {year} {2012})}\BibitemShut {NoStop}%
\bibitem [{\citenamefont {Esposito}\ \emph {et~al.}(2010)\citenamefont {Esposito}, \citenamefont {Kawai}, \citenamefont {Lindenberg},\ and\ \citenamefont {Van~den Broeck}}]{PhysRevLett.105.150603}%
  \BibitemOpen
  \bibfield  {author} {\bibinfo {author} {\bibfnamefont {M.}~\bibnamefont {Esposito}}, \bibinfo {author} {\bibfnamefont {R.}~\bibnamefont {Kawai}}, \bibinfo {author} {\bibfnamefont {K.}~\bibnamefont {Lindenberg}},\ and\ \bibinfo {author} {\bibfnamefont {C.}~\bibnamefont {Van~den Broeck}},\ }\bibfield  {title} {\bibinfo {title} {Efficiency at maximum power of low-dissipation carnot engines},\ }\href {https://doi.org/10.1103/PhysRevLett.105.150603} {\bibfield  {journal} {\bibinfo  {journal} {Phys. Rev. Lett.}\ }\textbf {\bibinfo {volume} {105}},\ \bibinfo {pages} {150603} (\bibinfo {year} {2010})}\BibitemShut {NoStop}%
\bibitem [{\citenamefont {Salamon}\ \emph {et~al.}(1980)\citenamefont {Salamon}, \citenamefont {Nitzan}, \citenamefont {Andresen},\ and\ \citenamefont {Berry}}]{PhysRevA.21.2115}%
  \BibitemOpen
  \bibfield  {author} {\bibinfo {author} {\bibfnamefont {P.}~\bibnamefont {Salamon}}, \bibinfo {author} {\bibfnamefont {A.}~\bibnamefont {Nitzan}}, \bibinfo {author} {\bibfnamefont {B.}~\bibnamefont {Andresen}},\ and\ \bibinfo {author} {\bibfnamefont {R.~S.}\ \bibnamefont {Berry}},\ }\bibfield  {title} {\bibinfo {title} {Minimum entropy production and the optimization of heat engines},\ }\href {https://doi.org/10.1103/PhysRevA.21.2115} {\bibfield  {journal} {\bibinfo  {journal} {Phys. Rev. A}\ }\textbf {\bibinfo {volume} {21}},\ \bibinfo {pages} {2115} (\bibinfo {year} {1980})}\BibitemShut {NoStop}%
\bibitem [{\citenamefont {Schmiedl}\ and\ \citenamefont {Seifert}(2007)}]{schmiedl2007efficiency}%
  \BibitemOpen
  \bibfield  {author} {\bibinfo {author} {\bibfnamefont {T.}~\bibnamefont {Schmiedl}}\ and\ \bibinfo {author} {\bibfnamefont {U.}~\bibnamefont {Seifert}},\ }\bibfield  {title} {\bibinfo {title} {Efficiency at maximum power: An analytically solvable model for stochastic heat engines},\ }\href@noop {} {\bibfield  {journal} {\bibinfo  {journal} {EPL}\ }\textbf {\bibinfo {volume} {81}},\ \bibinfo {pages} {20003} (\bibinfo {year} {2007})}\BibitemShut {NoStop}%
\bibitem [{\citenamefont {Curzon}\ and\ \citenamefont {Ahlborn}(1975)}]{curzon1975efficiency}%
  \BibitemOpen
  \bibfield  {author} {\bibinfo {author} {\bibfnamefont {F.~L.}\ \bibnamefont {Curzon}}\ and\ \bibinfo {author} {\bibfnamefont {B.}~\bibnamefont {Ahlborn}},\ }\bibfield  {title} {\bibinfo {title} {Efficiency of a carnot engine at maximum power output},\ }\href@noop {} {\bibfield  {journal} {\bibinfo  {journal} {Am. J. Phys.}\ }\textbf {\bibinfo {volume} {43}},\ \bibinfo {pages} {22} (\bibinfo {year} {1975})}\BibitemShut {NoStop}%
\bibitem [{\citenamefont {Chen}\ and\ \citenamefont {Yan}(1989)}]{Chen1989TheEO}%
  \BibitemOpen
  \bibfield  {author} {\bibinfo {author} {\bibfnamefont {L.}~\bibnamefont {Chen}}\ and\ \bibinfo {author} {\bibfnamefont {Z.}~\bibnamefont {Yan}},\ }\bibfield  {title} {\bibinfo {title} {The effect of heat‐transfer law on performance of a two‐heat‐source endoreversible cycle},\ }\href {https://api.semanticscholar.org/CorpusID:95868175} {\bibfield  {journal} {\bibinfo  {journal} {J. Chem. Phys.}\ }\textbf {\bibinfo {volume} {90}},\ \bibinfo {pages} {3740} (\bibinfo {year} {1989})}\BibitemShut {NoStop}%
\bibitem [{\citenamefont {De~Vos}(1985)}]{de1985efficiency}%
  \BibitemOpen
  \bibfield  {author} {\bibinfo {author} {\bibfnamefont {A.}~\bibnamefont {De~Vos}},\ }\bibfield  {title} {\bibinfo {title} {Efficiency of some heat engines at maximum-power conditions},\ }\href@noop {} {\bibfield  {journal} {\bibinfo  {journal} {Am. J. Phys.}\ }\textbf {\bibinfo {volume} {53}},\ \bibinfo {pages} {570} (\bibinfo {year} {1985})}\BibitemShut {NoStop}%
\bibitem [{\citenamefont {Van~den Broeck}(2005)}]{PhysRevLett.95.190602}%
  \BibitemOpen
  \bibfield  {author} {\bibinfo {author} {\bibfnamefont {C.}~\bibnamefont {Van~den Broeck}},\ }\bibfield  {title} {\bibinfo {title} {Thermodynamic efficiency at maximum power},\ }\href {https://doi.org/10.1103/PhysRevLett.95.190602} {\bibfield  {journal} {\bibinfo  {journal} {Phys. Rev. Lett.}\ }\textbf {\bibinfo {volume} {95}},\ \bibinfo {pages} {190602} (\bibinfo {year} {2005})}\BibitemShut {NoStop}%
\bibitem [{\citenamefont {Izumida}\ and\ \citenamefont {Okuda}(2014)}]{PhysRevLett.112.180603}%
  \BibitemOpen
  \bibfield  {author} {\bibinfo {author} {\bibfnamefont {Y.}~\bibnamefont {Izumida}}\ and\ \bibinfo {author} {\bibfnamefont {K.}~\bibnamefont {Okuda}},\ }\bibfield  {title} {\bibinfo {title} {Work output and efficiency at maximum power of linear irreversible heat engines operating with a finite-sized heat source},\ }\href {https://doi.org/10.1103/PhysRevLett.112.180603} {\bibfield  {journal} {\bibinfo  {journal} {Phys. Rev. Lett.}\ }\textbf {\bibinfo {volume} {112}},\ \bibinfo {pages} {180603} (\bibinfo {year} {2014})}\BibitemShut {NoStop}%
\bibitem [{\citenamefont {Jim\'enez~de Cisneros}\ and\ \citenamefont {Hern\'andez}(2008)}]{PhysRevE.77.041127}%
  \BibitemOpen
  \bibfield  {author} {\bibinfo {author} {\bibfnamefont {B.}~\bibnamefont {Jim\'enez~de Cisneros}}\ and\ \bibinfo {author} {\bibfnamefont {A.~C.}\ \bibnamefont {Hern\'andez}},\ }\bibfield  {title} {\bibinfo {title} {Coupled heat devices in linear irreversible thermodynamics},\ }\href {https://doi.org/10.1103/PhysRevE.77.041127} {\bibfield  {journal} {\bibinfo  {journal} {Phys. Rev. E}\ }\textbf {\bibinfo {volume} {77}},\ \bibinfo {pages} {041127} (\bibinfo {year} {2008})}\BibitemShut {NoStop}%
\bibitem [{\citenamefont {de~Tom\'as}\ \emph {et~al.}(2012)\citenamefont {de~Tom\'as}, \citenamefont {Hern\'andez},\ and\ \citenamefont {Roco}}]{PhysRevE.85.010104}%
  \BibitemOpen
  \bibfield  {author} {\bibinfo {author} {\bibfnamefont {C.}~\bibnamefont {de~Tom\'as}}, \bibinfo {author} {\bibfnamefont {A.~C.}\ \bibnamefont {Hern\'andez}},\ and\ \bibinfo {author} {\bibfnamefont {J.~M.~M.}\ \bibnamefont {Roco}},\ }\bibfield  {title} {\bibinfo {title} {Optimal low symmetric dissipation carnot engines and refrigerators},\ }\href {https://doi.org/10.1103/PhysRevE.85.010104} {\bibfield  {journal} {\bibinfo  {journal} {Phys. Rev. E}\ }\textbf {\bibinfo {volume} {85}},\ \bibinfo {pages} {010104} (\bibinfo {year} {2012})}\BibitemShut {NoStop}%
\bibitem [{\citenamefont {Ye}\ and\ \citenamefont {Holubec}(2021)}]{PhysRevE.103.052125}%
  \BibitemOpen
  \bibfield  {author} {\bibinfo {author} {\bibfnamefont {Z.}~\bibnamefont {Ye}}\ and\ \bibinfo {author} {\bibfnamefont {V.}~\bibnamefont {Holubec}},\ }\bibfield  {title} {\bibinfo {title} {Maximum efficiency of absorption refrigerators at arbitrary cooling power},\ }\href {https://doi.org/10.1103/PhysRevE.103.052125} {\bibfield  {journal} {\bibinfo  {journal} {Phys. Rev. E}\ }\textbf {\bibinfo {volume} {103}},\ \bibinfo {pages} {052125} (\bibinfo {year} {2021})}\BibitemShut {NoStop}%
\bibitem [{\citenamefont {Ye}\ and\ \citenamefont {Holubec}(2022)}]{PhysRevE.105.024139}%
  \BibitemOpen
  \bibfield  {author} {\bibinfo {author} {\bibfnamefont {Z.}~\bibnamefont {Ye}}\ and\ \bibinfo {author} {\bibfnamefont {V.}~\bibnamefont {Holubec}},\ }\bibfield  {title} {\bibinfo {title} {Maximum efficiency of low-dissipation heat pumps at given heating load},\ }\href {https://doi.org/10.1103/PhysRevE.105.024139} {\bibfield  {journal} {\bibinfo  {journal} {Phys. Rev. E}\ }\textbf {\bibinfo {volume} {105}},\ \bibinfo {pages} {024139} (\bibinfo {year} {2022})}\BibitemShut {NoStop}%
\bibitem [{\citenamefont {Abiuso}\ and\ \citenamefont {Perarnau-Llobet}(2020)}]{PhysRevLett.124.110606}%
  \BibitemOpen
  \bibfield  {author} {\bibinfo {author} {\bibfnamefont {P.}~\bibnamefont {Abiuso}}\ and\ \bibinfo {author} {\bibfnamefont {M.}~\bibnamefont {Perarnau-Llobet}},\ }\bibfield  {title} {\bibinfo {title} {Optimal cycles for low-dissipation heat engines},\ }\href {https://doi.org/10.1103/PhysRevLett.124.110606} {\bibfield  {journal} {\bibinfo  {journal} {Phys. Rev. Lett.}\ }\textbf {\bibinfo {volume} {124}},\ \bibinfo {pages} {110606} (\bibinfo {year} {2020})}\BibitemShut {NoStop}%
\bibitem [{\citenamefont {de~Tomas}\ \emph {et~al.}(2013)\citenamefont {de~Tomas}, \citenamefont {Roco}, \citenamefont {Hern\'andez}, \citenamefont {Wang},\ and\ \citenamefont {Tu}}]{PhysRevE.87.012105}%
  \BibitemOpen
  \bibfield  {author} {\bibinfo {author} {\bibfnamefont {C.}~\bibnamefont {de~Tomas}}, \bibinfo {author} {\bibfnamefont {J.~M.~M.}\ \bibnamefont {Roco}}, \bibinfo {author} {\bibfnamefont {A.~C.}\ \bibnamefont {Hern\'andez}}, \bibinfo {author} {\bibfnamefont {Y.}~\bibnamefont {Wang}},\ and\ \bibinfo {author} {\bibfnamefont {Z.~C.}\ \bibnamefont {Tu}},\ }\bibfield  {title} {\bibinfo {title} {Low-dissipation heat devices: Unified trade-off optimization and bounds},\ }\href {https://doi.org/10.1103/PhysRevE.87.012105} {\bibfield  {journal} {\bibinfo  {journal} {Phys. Rev. E}\ }\textbf {\bibinfo {volume} {87}},\ \bibinfo {pages} {012105} (\bibinfo {year} {2013})}\BibitemShut {NoStop}%
\bibitem [{\citenamefont {Reyes-Ram\'{\i}rez}\ \emph {et~al.}(2017)\citenamefont {Reyes-Ram\'{\i}rez}, \citenamefont {Gonzalez-Ayala}, \citenamefont {Calvo~Hern\'andez},\ and\ \citenamefont {Santill\'an}}]{PhysRevE.96.042128}%
  \BibitemOpen
  \bibfield  {author} {\bibinfo {author} {\bibfnamefont {I.}~\bibnamefont {Reyes-Ram\'{\i}rez}}, \bibinfo {author} {\bibfnamefont {J.}~\bibnamefont {Gonzalez-Ayala}}, \bibinfo {author} {\bibfnamefont {A.}~\bibnamefont {Calvo~Hern\'andez}},\ and\ \bibinfo {author} {\bibfnamefont {M.}~\bibnamefont {Santill\'an}},\ }\bibfield  {title} {\bibinfo {title} {Local-stability analysis of a low-dissipation heat engine working at maximum power output},\ }\href {https://doi.org/10.1103/PhysRevE.96.042128} {\bibfield  {journal} {\bibinfo  {journal} {Phys. Rev. E}\ }\textbf {\bibinfo {volume} {96}},\ \bibinfo {pages} {042128} (\bibinfo {year} {2017})}\BibitemShut {NoStop}%
\bibitem [{\citenamefont {Moukalled}\ \emph {et~al.}(1995)\citenamefont {Moukalled}, \citenamefont {Nuwayhid},\ and\ \citenamefont {Noueihed}}]{moukalled1995efficiency}%
  \BibitemOpen
  \bibfield  {author} {\bibinfo {author} {\bibfnamefont {F.}~\bibnamefont {Moukalled}}, \bibinfo {author} {\bibfnamefont {R.}~\bibnamefont {Nuwayhid}},\ and\ \bibinfo {author} {\bibfnamefont {N.}~\bibnamefont {Noueihed}},\ }\bibfield  {title} {\bibinfo {title} {The efficiency of endoreversible heat engines with heat leak},\ }\href@noop {} {\bibfield  {journal} {\bibinfo  {journal} {Int. J. Energy Res.}\ }\textbf {\bibinfo {volume} {19}},\ \bibinfo {pages} {377} (\bibinfo {year} {1995})}\BibitemShut {NoStop}%
\bibitem [{\citenamefont {Gordon}\ and\ \citenamefont {Huleihil}(1992)}]{gordon1992general}%
  \BibitemOpen
  \bibfield  {author} {\bibinfo {author} {\bibfnamefont {J.}~\bibnamefont {Gordon}}\ and\ \bibinfo {author} {\bibfnamefont {M.}~\bibnamefont {Huleihil}},\ }\bibfield  {title} {\bibinfo {title} {General performance characteristics of real heat engines},\ }\href@noop {} {\bibfield  {journal} {\bibinfo  {journal} {J. Appl. Phys.}\ }\textbf {\bibinfo {volume} {72}},\ \bibinfo {pages} {829} (\bibinfo {year} {1992})}\BibitemShut {NoStop}%
\bibitem [{\citenamefont {Chen}\ \emph {et~al.}(2002)\citenamefont {Chen}, \citenamefont {Zhou}, \citenamefont {Sun},\ and\ \citenamefont {Wu}}]{chen2002optimal}%
  \BibitemOpen
  \bibfield  {author} {\bibinfo {author} {\bibfnamefont {L.}~\bibnamefont {Chen}}, \bibinfo {author} {\bibfnamefont {S.}~\bibnamefont {Zhou}}, \bibinfo {author} {\bibfnamefont {F.}~\bibnamefont {Sun}},\ and\ \bibinfo {author} {\bibfnamefont {C.}~\bibnamefont {Wu}},\ }\bibfield  {title} {\bibinfo {title} {Optimal configuration and performance of heat engines with heat leak and finite heat capacity},\ }\href@noop {} {\bibfield  {journal} {\bibinfo  {journal} {Open Syst. Inf. Dyn.}\ }\textbf {\bibinfo {volume} {9}},\ \bibinfo {pages} {85} (\bibinfo {year} {2002})}\BibitemShut {NoStop}%
\bibitem [{\citenamefont {Wang}\ \emph {et~al.}(2015)\citenamefont {Wang}, \citenamefont {Lai}, \citenamefont {Ye}, \citenamefont {He}, \citenamefont {Ma},\ and\ \citenamefont {Liao}}]{PhysRevE.91.050102}%
  \BibitemOpen
  \bibfield  {author} {\bibinfo {author} {\bibfnamefont {J.}~\bibnamefont {Wang}}, \bibinfo {author} {\bibfnamefont {Y.}~\bibnamefont {Lai}}, \bibinfo {author} {\bibfnamefont {Z.}~\bibnamefont {Ye}}, \bibinfo {author} {\bibfnamefont {J.}~\bibnamefont {He}}, \bibinfo {author} {\bibfnamefont {Y.}~\bibnamefont {Ma}},\ and\ \bibinfo {author} {\bibfnamefont {Q.}~\bibnamefont {Liao}},\ }\bibfield  {title} {\bibinfo {title} {Four-level refrigerator driven by photons},\ }\href {https://doi.org/10.1103/PhysRevE.91.050102} {\bibfield  {journal} {\bibinfo  {journal} {Phys. Rev. E}\ }\textbf {\bibinfo {volume} {91}},\ \bibinfo {pages} {050102} (\bibinfo {year} {2015})}\BibitemShut {NoStop}%
\bibitem [{\citenamefont {Bauer}\ \emph {et~al.}(2016)\citenamefont {Bauer}, \citenamefont {Brandner},\ and\ \citenamefont {Seifert}}]{PhysRevE.93.042112}%
  \BibitemOpen
  \bibfield  {author} {\bibinfo {author} {\bibfnamefont {M.}~\bibnamefont {Bauer}}, \bibinfo {author} {\bibfnamefont {K.}~\bibnamefont {Brandner}},\ and\ \bibinfo {author} {\bibfnamefont {U.}~\bibnamefont {Seifert}},\ }\bibfield  {title} {\bibinfo {title} {Optimal performance of periodically driven, stochastic heat engines under limited control},\ }\href {https://doi.org/10.1103/PhysRevE.93.042112} {\bibfield  {journal} {\bibinfo  {journal} {Phys. Rev. E}\ }\textbf {\bibinfo {volume} {93}},\ \bibinfo {pages} {042112} (\bibinfo {year} {2016})}\BibitemShut {NoStop}%
\bibitem [{\citenamefont {Chen}(1994)}]{chen1994maximum}%
  \BibitemOpen
  \bibfield  {author} {\bibinfo {author} {\bibfnamefont {J.}~\bibnamefont {Chen}},\ }\bibfield  {title} {\bibinfo {title} {The maximum power output and maximum efficiency of an irreversible carnot heat engine},\ }\href@noop {} {\bibfield  {journal} {\bibinfo  {journal} {J. Phys. D Appl. Phys.}\ }\textbf {\bibinfo {volume} {27}},\ \bibinfo {pages} {1144} (\bibinfo {year} {1994})}\BibitemShut {NoStop}%
\bibitem [{\citenamefont {Izumida}\ and\ \citenamefont {Okuda}(2012)}]{izumida2012efficiency}%
  \BibitemOpen
  \bibfield  {author} {\bibinfo {author} {\bibfnamefont {Y.}~\bibnamefont {Izumida}}\ and\ \bibinfo {author} {\bibfnamefont {K.}~\bibnamefont {Okuda}},\ }\bibfield  {title} {\bibinfo {title} {Efficiency at maximum power of minimally nonlinear irreversible heat engines},\ }\href@noop {} {\bibfield  {journal} {\bibinfo  {journal} {EPL}\ }\textbf {\bibinfo {volume} {97}},\ \bibinfo {pages} {10004} (\bibinfo {year} {2012})}\BibitemShut {NoStop}%
\bibitem [{\citenamefont {Izumida}\ \emph {et~al.}(2015)\citenamefont {Izumida}, \citenamefont {Okuda}, \citenamefont {Roco},\ and\ \citenamefont {Hern\'andez}}]{PhysRevE.91.052140}%
  \BibitemOpen
  \bibfield  {author} {\bibinfo {author} {\bibfnamefont {Y.}~\bibnamefont {Izumida}}, \bibinfo {author} {\bibfnamefont {K.}~\bibnamefont {Okuda}}, \bibinfo {author} {\bibfnamefont {J.~M.~M.}\ \bibnamefont {Roco}},\ and\ \bibinfo {author} {\bibfnamefont {A.~C.}\ \bibnamefont {Hern\'andez}},\ }\bibfield  {title} {\bibinfo {title} {Heat devices in nonlinear irreversible thermodynamics},\ }\href {https://doi.org/10.1103/PhysRevE.91.052140} {\bibfield  {journal} {\bibinfo  {journal} {Phys. Rev. E}\ }\textbf {\bibinfo {volume} {91}},\ \bibinfo {pages} {052140} (\bibinfo {year} {2015})}\BibitemShut {NoStop}%
\bibitem [{\citenamefont {Holubec}\ and\ \citenamefont {Ye}(2020)}]{PhysRevE.101.052124}%
  \BibitemOpen
  \bibfield  {author} {\bibinfo {author} {\bibfnamefont {V.}~\bibnamefont {Holubec}}\ and\ \bibinfo {author} {\bibfnamefont {Z.}~\bibnamefont {Ye}},\ }\bibfield  {title} {\bibinfo {title} {Maximum efficiency of low-dissipation refrigerators at arbitrary cooling power},\ }\href {https://doi.org/10.1103/PhysRevE.101.052124} {\bibfield  {journal} {\bibinfo  {journal} {Phys. Rev. E}\ }\textbf {\bibinfo {volume} {101}},\ \bibinfo {pages} {052124} (\bibinfo {year} {2020})}\BibitemShut {NoStop}%
\bibitem [{\citenamefont {Zhang}\ and\ \citenamefont {He}(2014)}]{50influence}%
  \BibitemOpen
  \bibfield  {author} {\bibinfo {author} {\bibfnamefont {Y.}~\bibnamefont {Zhang}}\ and\ \bibinfo {author} {\bibfnamefont {J.}~\bibnamefont {He}},\ }\bibfield  {title} {\bibinfo {title} {Influence of heat leak on efficiency at maximum power and its bounds of low-dissipation carnot heat engine},\ }\href@noop {} {\bibfield  {journal} {\bibinfo  {journal} {J. Mech. Eng.}\ }\textbf {\bibinfo {volume} {50}},\ \bibinfo {pages} {163} (\bibinfo {year} {2014})}\BibitemShut {NoStop}%
\bibitem [{\citenamefont {Huang}\ \emph {et~al.}(2015)\citenamefont {Huang}, \citenamefont {Guo},\ and\ \citenamefont {Chen}}]{huang2015performance}%
  \BibitemOpen
  \bibfield  {author} {\bibinfo {author} {\bibfnamefont {C.-K.}\ \bibnamefont {Huang}}, \bibinfo {author} {\bibfnamefont {J.-C.}\ \bibnamefont {Guo}},\ and\ \bibinfo {author} {\bibfnamefont {J.-C.}\ \bibnamefont {Chen}},\ }\bibfield  {title} {\bibinfo {title} {Performance characteristics of low-dissipative generalized carnot cycles with external leakage losses},\ }\href@noop {} {\bibfield  {journal} {\bibinfo  {journal} {Chin. Phys. B}\ }\textbf {\bibinfo {volume} {24}},\ \bibinfo {pages} {110506} (\bibinfo {year} {2015})}\BibitemShut {NoStop}%
\bibitem [{\citenamefont {Reif}(2009)}]{reif2009fundamentals}%
  \BibitemOpen
  \bibfield  {author} {\bibinfo {author} {\bibfnamefont {F.}~\bibnamefont {Reif}},\ }\href@noop {} {\emph {\bibinfo {title} {Fundamentals of statistical and thermal physics}}}\ (\bibinfo  {publisher} {Waveland Press},\ \bibinfo {year} {2009})\BibitemShut {NoStop}%
\bibitem [{\citenamefont {Bejan}(2016)}]{bejan2016advanced}%
  \BibitemOpen
  \bibfield  {author} {\bibinfo {author} {\bibfnamefont {A.}~\bibnamefont {Bejan}},\ }\href@noop {} {\emph {\bibinfo {title} {Advanced engineering thermodynamics}}}\ (\bibinfo  {publisher} {John Wiley \& Sons},\ \bibinfo {year} {2016})\BibitemShut {NoStop}%
\bibitem [{\citenamefont {Dittman}\ and\ \citenamefont {Zemansky}(1997)}]{dittman2021heat}%
  \BibitemOpen
  \bibfield  {author} {\bibinfo {author} {\bibfnamefont {R.~H.}\ \bibnamefont {Dittman}}\ and\ \bibinfo {author} {\bibfnamefont {M.~W.}\ \bibnamefont {Zemansky}},\ }\href@noop {} {\emph {\bibinfo {title} {Heat and thermodynamics, 7th ed.}}}\ (\bibinfo  {publisher} {McGraw-Hill},\ \bibinfo {year} {1997})\BibitemShut {NoStop}%
\bibitem [{\citenamefont {Chen}\ \emph {et~al.}(1997)\citenamefont {Chen}, \citenamefont {Sun},\ and\ \citenamefont {Wu}}]{chen1997influence}%
  \BibitemOpen
  \bibfield  {author} {\bibinfo {author} {\bibfnamefont {L.}~\bibnamefont {Chen}}, \bibinfo {author} {\bibfnamefont {F.}~\bibnamefont {Sun}},\ and\ \bibinfo {author} {\bibfnamefont {C.}~\bibnamefont {Wu}},\ }\bibfield  {title} {\bibinfo {title} {Influence of internal heat leak on the power versus efficiency characteristics of heat engines},\ }\href@noop {} {\bibfield  {journal} {\bibinfo  {journal} {Energy Convers. Manag.}\ }\textbf {\bibinfo {volume} {38}},\ \bibinfo {pages} {1501} (\bibinfo {year} {1997})}\BibitemShut {NoStop}%
\bibitem [{\citenamefont {Izumida}\ \emph {et~al.}(2013)\citenamefont {Izumida}, \citenamefont {Okuda}, \citenamefont {Hern{\'a}ndez},\ and\ \citenamefont {Roco}}]{izumida2013coefficient}%
  \BibitemOpen
  \bibfield  {author} {\bibinfo {author} {\bibfnamefont {Y.}~\bibnamefont {Izumida}}, \bibinfo {author} {\bibfnamefont {K.}~\bibnamefont {Okuda}}, \bibinfo {author} {\bibfnamefont {A.~C.}\ \bibnamefont {Hern{\'a}ndez}},\ and\ \bibinfo {author} {\bibfnamefont {J.}~\bibnamefont {Roco}},\ }\bibfield  {title} {\bibinfo {title} {Coefficient of performance under optimized figure of merit in minimally nonlinear irreversible refrigerator},\ }\href@noop {} {\bibfield  {journal} {\bibinfo  {journal} {EPL}\ }\textbf {\bibinfo {volume} {101}},\ \bibinfo {pages} {10005} (\bibinfo {year} {2013})}\BibitemShut {NoStop}%
\bibitem [{\citenamefont {Guo}\ \emph {et~al.}(2019)\citenamefont {Guo}, \citenamefont {Yang}, \citenamefont {Zhang}, \citenamefont {Gonzalez-Ayala}, \citenamefont {Roco}, \citenamefont {Medina},\ and\ \citenamefont {Hern{\'a}ndez}}]{guo2019thermally}%
  \BibitemOpen
  \bibfield  {author} {\bibinfo {author} {\bibfnamefont {J.}~\bibnamefont {Guo}}, \bibinfo {author} {\bibfnamefont {H.}~\bibnamefont {Yang}}, \bibinfo {author} {\bibfnamefont {H.}~\bibnamefont {Zhang}}, \bibinfo {author} {\bibfnamefont {J.}~\bibnamefont {Gonzalez-Ayala}}, \bibinfo {author} {\bibfnamefont {J.}~\bibnamefont {Roco}}, \bibinfo {author} {\bibfnamefont {A.}~\bibnamefont {Medina}},\ and\ \bibinfo {author} {\bibfnamefont {A.~C.}\ \bibnamefont {Hern{\'a}ndez}},\ }\bibfield  {title} {\bibinfo {title} {Thermally driven refrigerators: Equivalent low-dissipation three-heat-source model and comparison with experimental and simulated results},\ }\href@noop {} {\bibfield  {journal} {\bibinfo  {journal} {Energy Convers. Manag.}\ }\textbf {\bibinfo {volume} {198}},\ \bibinfo {pages} {111917} (\bibinfo {year} {2019})}\BibitemShut {NoStop}%
\bibitem [{\citenamefont {Ahmadi}\ \emph {et~al.}(2015)\citenamefont {Ahmadi}, \citenamefont {Ahmadi}, \citenamefont {Mehrpooya},\ and\ \citenamefont {Sameti}}]{ahmadi2015thermo}%
  \BibitemOpen
  \bibfield  {author} {\bibinfo {author} {\bibfnamefont {M.~H.}\ \bibnamefont {Ahmadi}}, \bibinfo {author} {\bibfnamefont {M.~A.}\ \bibnamefont {Ahmadi}}, \bibinfo {author} {\bibfnamefont {M.}~\bibnamefont {Mehrpooya}},\ and\ \bibinfo {author} {\bibfnamefont {M.}~\bibnamefont {Sameti}},\ }\bibfield  {title} {\bibinfo {title} {Thermo-ecological analysis and optimization performance of an irreversible three-heat-source absorption heat pump},\ }\href@noop {} {\bibfield  {journal} {\bibinfo  {journal} {Energy Convers. Manag.}\ }\textbf {\bibinfo {volume} {90}},\ \bibinfo {pages} {175} (\bibinfo {year} {2015})}\BibitemShut {NoStop}%
\bibitem [{\citenamefont {Holubec}\ and\ \citenamefont {Ryabov}(2017)}]{PhysRevE.96.062107}%
  \BibitemOpen
  \bibfield  {author} {\bibinfo {author} {\bibfnamefont {V.}~\bibnamefont {Holubec}}\ and\ \bibinfo {author} {\bibfnamefont {A.}~\bibnamefont {Ryabov}},\ }\bibfield  {title} {\bibinfo {title} {Diverging, but negligible power at carnot efficiency: Theory and experiment},\ }\href {https://doi.org/10.1103/PhysRevE.96.062107} {\bibfield  {journal} {\bibinfo  {journal} {Phys. Rev. E}\ }\textbf {\bibinfo {volume} {96}},\ \bibinfo {pages} {062107} (\bibinfo {year} {2017})}\BibitemShut {NoStop}%
\bibitem [{\citenamefont {Iyyappan}\ and\ \citenamefont {Johal}(2020)}]{iyyappan2020efficiency}%
  \BibitemOpen
  \bibfield  {author} {\bibinfo {author} {\bibfnamefont {I.}~\bibnamefont {Iyyappan}}\ and\ \bibinfo {author} {\bibfnamefont {R.~S.}\ \bibnamefont {Johal}},\ }\bibfield  {title} {\bibinfo {title} {Efficiency of a two-stage heat engine at optimal power},\ }\href@noop {} {\bibfield  {journal} {\bibinfo  {journal} {EPL}\ }\textbf {\bibinfo {volume} {128}},\ \bibinfo {pages} {50004} (\bibinfo {year} {2020})}\BibitemShut {NoStop}%
\bibitem [{\citenamefont {Zulkowski}\ and\ \citenamefont {DeWeese}(2015)}]{PhysRevE.92.032113}%
  \BibitemOpen
  \bibfield  {author} {\bibinfo {author} {\bibfnamefont {P.~R.}\ \bibnamefont {Zulkowski}}\ and\ \bibinfo {author} {\bibfnamefont {M.~R.}\ \bibnamefont {DeWeese}},\ }\bibfield  {title} {\bibinfo {title} {Optimal protocols for slowly driven quantum systems},\ }\href {https://doi.org/10.1103/PhysRevE.92.032113} {\bibfield  {journal} {\bibinfo  {journal} {Phys. Rev. E}\ }\textbf {\bibinfo {volume} {92}},\ \bibinfo {pages} {032113} (\bibinfo {year} {2015})}\BibitemShut {NoStop}%
\bibitem [{\citenamefont {Zhao}\ \emph {et~al.}(2022)\citenamefont {Zhao}, \citenamefont {Gong},\ and\ \citenamefont {Tu}}]{zhao2022microscopic}%
  \BibitemOpen
  \bibfield  {author} {\bibinfo {author} {\bibfnamefont {X.-H.}\ \bibnamefont {Zhao}}, \bibinfo {author} {\bibfnamefont {Z.-N.}\ \bibnamefont {Gong}},\ and\ \bibinfo {author} {\bibfnamefont {Z.~C.}\ \bibnamefont {Tu}},\ }\bibfield  {title} {\bibinfo {title} {Low-dissipation engines: Microscopic construction via shortcuts to adiabaticity and isothermality, the optimal relation between power and efficiency},\ }\href {https://doi.org/10.1103/PhysRevE.106.064117} {\bibfield  {journal} {\bibinfo  {journal} {Phys. Rev. E}\ }\textbf {\bibinfo {volume} {106}},\ \bibinfo {pages} {064117} (\bibinfo {year} {2022})}\BibitemShut {NoStop}%
\bibitem [{\citenamefont {Mart{\'\i}nez}\ \emph {et~al.}(2016)\citenamefont {Mart{\'\i}nez}, \citenamefont {Rold{\'a}n}, \citenamefont {Dinis}, \citenamefont {Petrov}, \citenamefont {Parrondo},\ and\ \citenamefont {Rica}}]{martinez2016brownian}%
  \BibitemOpen
  \bibfield  {author} {\bibinfo {author} {\bibfnamefont {I.~A.}\ \bibnamefont {Mart{\'\i}nez}}, \bibinfo {author} {\bibfnamefont {{\'E}.}~\bibnamefont {Rold{\'a}n}}, \bibinfo {author} {\bibfnamefont {L.}~\bibnamefont {Dinis}}, \bibinfo {author} {\bibfnamefont {D.}~\bibnamefont {Petrov}}, \bibinfo {author} {\bibfnamefont {J.~M.}\ \bibnamefont {Parrondo}},\ and\ \bibinfo {author} {\bibfnamefont {R.~A.}\ \bibnamefont {Rica}},\ }\bibfield  {title} {\bibinfo {title} {Brownian carnot engine},\ }\href@noop {} {\bibfield  {journal} {\bibinfo  {journal} {Nat. Phys.}\ }\textbf {\bibinfo {volume} {12}},\ \bibinfo {pages} {67} (\bibinfo {year} {2016})}\BibitemShut {NoStop}%
\bibitem [{\citenamefont {Gonzalez-Ayala}\ \emph {et~al.}(2018)\citenamefont {Gonzalez-Ayala}, \citenamefont {Medina}, \citenamefont {Roco},\ and\ \citenamefont {Hern\'andez}}]{PhysRevE.97.022139}%
  \BibitemOpen
  \bibfield  {author} {\bibinfo {author} {\bibfnamefont {J.}~\bibnamefont {Gonzalez-Ayala}}, \bibinfo {author} {\bibfnamefont {A.}~\bibnamefont {Medina}}, \bibinfo {author} {\bibfnamefont {J.~M.~M.}\ \bibnamefont {Roco}},\ and\ \bibinfo {author} {\bibfnamefont {A.~C.}\ \bibnamefont {Hern\'andez}},\ }\bibfield  {title} {\bibinfo {title} {Entropy generation and unified optimization of carnot-like and low-dissipation refrigerators},\ }\href {https://doi.org/10.1103/PhysRevE.97.022139} {\bibfield  {journal} {\bibinfo  {journal} {Phys. Rev. E}\ }\textbf {\bibinfo {volume} {97}},\ \bibinfo {pages} {022139} (\bibinfo {year} {2018})}\BibitemShut {NoStop}%
\bibitem [{\citenamefont {Dechant}\ \emph {et~al.}(2017)\citenamefont {Dechant}, \citenamefont {Kiesel},\ and\ \citenamefont {Lutz}}]{dechant2017underdamped}%
  \BibitemOpen
  \bibfield  {author} {\bibinfo {author} {\bibfnamefont {A.}~\bibnamefont {Dechant}}, \bibinfo {author} {\bibfnamefont {N.}~\bibnamefont {Kiesel}},\ and\ \bibinfo {author} {\bibfnamefont {E.}~\bibnamefont {Lutz}},\ }\bibfield  {title} {\bibinfo {title} {Underdamped stochastic heat engine at maximum efficiency},\ }\href@noop {} {\bibfield  {journal} {\bibinfo  {journal} {EPL}\ }\textbf {\bibinfo {volume} {119}},\ \bibinfo {pages} {50003} (\bibinfo {year} {2017})}\BibitemShut {NoStop}%
\bibitem [{\citenamefont {Ye}\ \emph {et~al.}(2022)\citenamefont {Ye}, \citenamefont {Cerisola}, \citenamefont {Abiuso}, \citenamefont {Anders}, \citenamefont {Perarnau-Llobet},\ and\ \citenamefont {Holubec}}]{ye2022optimal}%
  \BibitemOpen
  \bibfield  {author} {\bibinfo {author} {\bibfnamefont {Z.}~\bibnamefont {Ye}}, \bibinfo {author} {\bibfnamefont {F.}~\bibnamefont {Cerisola}}, \bibinfo {author} {\bibfnamefont {P.}~\bibnamefont {Abiuso}}, \bibinfo {author} {\bibfnamefont {J.}~\bibnamefont {Anders}}, \bibinfo {author} {\bibfnamefont {M.}~\bibnamefont {Perarnau-Llobet}},\ and\ \bibinfo {author} {\bibfnamefont {V.}~\bibnamefont {Holubec}},\ }\bibfield  {title} {\bibinfo {title} {Optimal finite-time heat engines under constrained control},\ }\href {https://doi.org/10.1103/PhysRevResearch.4.043130} {\bibfield  {journal} {\bibinfo  {journal} {Phys. Rev. Res.}\ }\textbf {\bibinfo {volume} {4}},\ \bibinfo {pages} {043130} (\bibinfo {year} {2022})}\BibitemShut {NoStop}%
\bibitem [{\citenamefont {Yan}\ and\ \citenamefont {Chen}(1990)}]{yan1990class}%
  \BibitemOpen
  \bibfield  {author} {\bibinfo {author} {\bibfnamefont {Z.}~\bibnamefont {Yan}}\ and\ \bibinfo {author} {\bibfnamefont {J.}~\bibnamefont {Chen}},\ }\bibfield  {title} {\bibinfo {title} {A class of irreversible carnot refrigeration cycles with a general heat transfer law},\ }\href@noop {} {\bibfield  {journal} {\bibinfo  {journal} {J. Phys. D Appl. Phys.}\ }\textbf {\bibinfo {volume} {23}},\ \bibinfo {pages} {136} (\bibinfo {year} {1990})}\BibitemShut {NoStop}%
\bibitem [{\citenamefont {Hu}\ \emph {et~al.}(2013)\citenamefont {Hu}, \citenamefont {Wu}, \citenamefont {Ma}, \citenamefont {He}, \citenamefont {Wang}, \citenamefont {Hern\'andez},\ and\ \citenamefont {Roco}}]{PhysRevE.88.062115}%
  \BibitemOpen
  \bibfield  {author} {\bibinfo {author} {\bibfnamefont {Y.}~\bibnamefont {Hu}}, \bibinfo {author} {\bibfnamefont {F.}~\bibnamefont {Wu}}, \bibinfo {author} {\bibfnamefont {Y.}~\bibnamefont {Ma}}, \bibinfo {author} {\bibfnamefont {J.}~\bibnamefont {He}}, \bibinfo {author} {\bibfnamefont {J.}~\bibnamefont {Wang}}, \bibinfo {author} {\bibfnamefont {A.~C.}\ \bibnamefont {Hern\'andez}},\ and\ \bibinfo {author} {\bibfnamefont {J.~M.~M.}\ \bibnamefont {Roco}},\ }\bibfield  {title} {\bibinfo {title} {Coefficient of performance for a low-dissipation carnot-like refrigerator with nonadiabatic dissipation},\ }\href {https://doi.org/10.1103/PhysRevE.88.062115} {\bibfield  {journal} {\bibinfo  {journal} {Phys. Rev. E}\ }\textbf {\bibinfo {volume} {88}},\ \bibinfo {pages} {062115} (\bibinfo {year} {2013})}\BibitemShut {NoStop}%
\bibitem [{\citenamefont {Holubec}\ and\ \citenamefont {Ryabov}(2016)}]{holubec2016maximum}%
  \BibitemOpen
  \bibfield  {author} {\bibinfo {author} {\bibfnamefont {V.}~\bibnamefont {Holubec}}\ and\ \bibinfo {author} {\bibfnamefont {A.}~\bibnamefont {Ryabov}},\ }\bibfield  {title} {\bibinfo {title} {Maximum efficiency of low-dissipation heat engines at arbitrary power},\ }\href@noop {} {\bibfield  {journal} {\bibinfo  {journal} {J. Stat. Mech. Theor. Exp.}\ }\textbf {\bibinfo {volume} {2016}},\ \bibinfo {pages} {073204} (\bibinfo {year} {2016})}\BibitemShut {NoStop}%
\bibitem [{\citenamefont {Blickle}\ and\ \citenamefont {Bechinger}(2012)}]{blickle2012realization}%
  \BibitemOpen
  \bibfield  {author} {\bibinfo {author} {\bibfnamefont {V.}~\bibnamefont {Blickle}}\ and\ \bibinfo {author} {\bibfnamefont {C.}~\bibnamefont {Bechinger}},\ }\bibfield  {title} {\bibinfo {title} {Realization of a micrometre-sized stochastic heat engine},\ }\href@noop {} {\bibfield  {journal} {\bibinfo  {journal} {Nat. Phys.}\ }\textbf {\bibinfo {volume} {8}},\ \bibinfo {pages} {143} (\bibinfo {year} {2012})}\BibitemShut {NoStop}%
\bibitem [{\citenamefont {Ma}\ \emph {et~al.}(2018)\citenamefont {Ma}, \citenamefont {Xu}, \citenamefont {Dong},\ and\ \citenamefont {Sun}}]{PhysRevE.98.042112}%
  \BibitemOpen
  \bibfield  {author} {\bibinfo {author} {\bibfnamefont {Y.-H.}\ \bibnamefont {Ma}}, \bibinfo {author} {\bibfnamefont {D.}~\bibnamefont {Xu}}, \bibinfo {author} {\bibfnamefont {H.}~\bibnamefont {Dong}},\ and\ \bibinfo {author} {\bibfnamefont {C.-P.}\ \bibnamefont {Sun}},\ }\bibfield  {title} {\bibinfo {title} {Universal constraint for efficiency and power of a low-dissipation heat engine},\ }\href {https://doi.org/10.1103/PhysRevE.98.042112} {\bibfield  {journal} {\bibinfo  {journal} {Phys. Rev. E}\ }\textbf {\bibinfo {volume} {98}},\ \bibinfo {pages} {042112} (\bibinfo {year} {2018})}\BibitemShut {NoStop}%
\bibitem [{\citenamefont {Holubec}\ and\ \citenamefont {Ryabov}(2015)}]{PhysRevE.92.052125}%
  \BibitemOpen
  \bibfield  {author} {\bibinfo {author} {\bibfnamefont {V.}~\bibnamefont {Holubec}}\ and\ \bibinfo {author} {\bibfnamefont {A.}~\bibnamefont {Ryabov}},\ }\bibfield  {title} {\bibinfo {title} {Efficiency at and near maximum power of low-dissipation heat engines},\ }\href {https://doi.org/10.1103/PhysRevE.92.052125} {\bibfield  {journal} {\bibinfo  {journal} {Phys. Rev. E}\ }\textbf {\bibinfo {volume} {92}},\ \bibinfo {pages} {052125} (\bibinfo {year} {2015})}\BibitemShut {NoStop}%
\bibitem [{\citenamefont {Esposito}\ \emph {et~al.}(2009)\citenamefont {Esposito}, \citenamefont {Lindenberg},\ and\ \citenamefont {Van~den Broeck}}]{PhysRevLett.102.130602}%
  \BibitemOpen
  \bibfield  {author} {\bibinfo {author} {\bibfnamefont {M.}~\bibnamefont {Esposito}}, \bibinfo {author} {\bibfnamefont {K.}~\bibnamefont {Lindenberg}},\ and\ \bibinfo {author} {\bibfnamefont {C.}~\bibnamefont {Van~den Broeck}},\ }\bibfield  {title} {\bibinfo {title} {Universality of efficiency at maximum power},\ }\href {https://doi.org/10.1103/PhysRevLett.102.130602} {\bibfield  {journal} {\bibinfo  {journal} {Phys. Rev. Lett.}\ }\textbf {\bibinfo {volume} {102}},\ \bibinfo {pages} {130602} (\bibinfo {year} {2009})}\BibitemShut {NoStop}%
\bibitem [{\citenamefont {Hern{\'a}ndez}\ \emph {et~al.}(2015)\citenamefont {Hern{\'a}ndez}, \citenamefont {Medina},\ and\ \citenamefont {Roco}}]{hernandez2015time}%
  \BibitemOpen
  \bibfield  {author} {\bibinfo {author} {\bibfnamefont {A.~C.}\ \bibnamefont {Hern{\'a}ndez}}, \bibinfo {author} {\bibfnamefont {A.}~\bibnamefont {Medina}},\ and\ \bibinfo {author} {\bibfnamefont {J.}~\bibnamefont {Roco}},\ }\bibfield  {title} {\bibinfo {title} {Time, entropy generation, and optimization in low-dissipation heat devices},\ }\href@noop {} {\bibfield  {journal} {\bibinfo  {journal} {New J. Phys.}\ }\textbf {\bibinfo {volume} {17}},\ \bibinfo {pages} {075011} (\bibinfo {year} {2015})}\BibitemShut {NoStop}%
\bibitem [{\citenamefont {Chen}\ \emph {et~al.}(1995)\citenamefont {Chen}, \citenamefont {Sun}, \citenamefont {Cheng},\ and\ \citenamefont {Chen}}]{chen1995study}%
  \BibitemOpen
  \bibfield  {author} {\bibinfo {author} {\bibfnamefont {W.}~\bibnamefont {Chen}}, \bibinfo {author} {\bibfnamefont {F.}~\bibnamefont {Sun}}, \bibinfo {author} {\bibfnamefont {S.}~\bibnamefont {Cheng}},\ and\ \bibinfo {author} {\bibfnamefont {L.}~\bibnamefont {Chen}},\ }\bibfield  {title} {\bibinfo {title} {Study on optimal performance and working temperatures of endoreversible forward and reverse carnot cycles},\ }\href@noop {} {\bibfield  {journal} {\bibinfo  {journal} {Int. J. Energy Res.}\ }\textbf {\bibinfo {volume} {19}},\ \bibinfo {pages} {751} (\bibinfo {year} {1995})}\BibitemShut {NoStop}%
\bibitem [{\citenamefont {Chen}\ \emph {et~al.}(2022)\citenamefont {Chen}, \citenamefont {Chen}, \citenamefont {Fei},\ and\ \citenamefont {Quan}}]{chen2021microscopic}%
  \BibitemOpen
  \bibfield  {author} {\bibinfo {author} {\bibfnamefont {Y.~H.}\ \bibnamefont {Chen}}, \bibinfo {author} {\bibfnamefont {J.-F.}\ \bibnamefont {Chen}}, \bibinfo {author} {\bibfnamefont {Z.}~\bibnamefont {Fei}},\ and\ \bibinfo {author} {\bibfnamefont {H.~T.}\ \bibnamefont {Quan}},\ }\bibfield  {title} {\bibinfo {title} {Microscopic theory of the curzon-ahlborn heat engine based on a brownian particle},\ }\href {https://doi.org/10.1103/PhysRevE.106.024105} {\bibfield  {journal} {\bibinfo  {journal} {Phys. Rev. E}\ }\textbf {\bibinfo {volume} {106}},\ \bibinfo {pages} {024105} (\bibinfo {year} {2022})}\BibitemShut {NoStop}%
\bibitem [{\citenamefont {Ye}\ \emph {et~al.}(2017)\citenamefont {Ye}, \citenamefont {Hu}, \citenamefont {He},\ and\ \citenamefont {Wang}}]{ye2017universality}%
  \BibitemOpen
  \bibfield  {author} {\bibinfo {author} {\bibfnamefont {Z.}~\bibnamefont {Ye}}, \bibinfo {author} {\bibfnamefont {Y.}~\bibnamefont {Hu}}, \bibinfo {author} {\bibfnamefont {J.}~\bibnamefont {He}},\ and\ \bibinfo {author} {\bibfnamefont {J.}~\bibnamefont {Wang}},\ }\bibfield  {title} {\bibinfo {title} {Universality of maximum-work efficiency of a cyclic heat engine based on a finite system of ultracold atoms},\ }\href@noop {} {\bibfield  {journal} {\bibinfo  {journal} {Sci. Rep.}\ }\textbf {\bibinfo {volume} {7}},\ \bibinfo {pages} {6289} (\bibinfo {year} {2017})}\BibitemShut {NoStop}%
\bibitem [{\citenamefont {Chen}(2022)}]{PhysRevE.106.054108}%
  \BibitemOpen
  \bibfield  {author} {\bibinfo {author} {\bibfnamefont {J.-F.}\ \bibnamefont {Chen}},\ }\bibfield  {title} {\bibinfo {title} {Optimizing brownian heat engine with shortcut strategy},\ }\href {https://doi.org/10.1103/PhysRevE.106.054108} {\bibfield  {journal} {\bibinfo  {journal} {Phys. Rev. E}\ }\textbf {\bibinfo {volume} {106}},\ \bibinfo {pages} {054108} (\bibinfo {year} {2022})}\BibitemShut {NoStop}%
\bibitem [{\citenamefont {Abiuso}\ \emph {et~al.}(2020)\citenamefont {Abiuso}, \citenamefont {Miller}, \citenamefont {Perarnau-Llobet},\ and\ \citenamefont {Scandi}}]{abiuso2020geometric}%
  \BibitemOpen
  \bibfield  {author} {\bibinfo {author} {\bibfnamefont {P.}~\bibnamefont {Abiuso}}, \bibinfo {author} {\bibfnamefont {H.~J.}\ \bibnamefont {Miller}}, \bibinfo {author} {\bibfnamefont {M.}~\bibnamefont {Perarnau-Llobet}},\ and\ \bibinfo {author} {\bibfnamefont {M.}~\bibnamefont {Scandi}},\ }\bibfield  {title} {\bibinfo {title} {Geometric optimisation of quantum thermodynamic processes},\ }\href@noop {} {\bibfield  {journal} {\bibinfo  {journal} {Entropy}\ }\textbf {\bibinfo {volume} {22}},\ \bibinfo {pages} {1076} (\bibinfo {year} {2020})}\BibitemShut {NoStop}%
\bibitem [{\citenamefont {Ma}\ and\ \citenamefont {Fu}(2024)}]{ma2024unified}%
  \BibitemOpen
  \bibfield  {author} {\bibinfo {author} {\bibfnamefont {Y.-H.}\ \bibnamefont {Ma}}\ and\ \bibinfo {author} {\bibfnamefont {C.}~\bibnamefont {Fu}},\ }\bibfield  {title} {\bibinfo {title} {Unified approach to power-efficiency trade-off of generic thermal machines},\ }\href@noop {} {\bibfield  {journal} {\bibinfo  {journal} {arXiv preprint arXiv:2411.03849}\ } (\bibinfo {year} {2024})}\BibitemShut {NoStop}%
\bibitem [{\citenamefont {Chen}\ \emph {et~al.}(2001)\citenamefont {Chen}, \citenamefont {Yan}, \citenamefont {Lin},\ and\ \citenamefont {Andresen}}]{chen2001curzon}%
  \BibitemOpen
  \bibfield  {author} {\bibinfo {author} {\bibfnamefont {J.}~\bibnamefont {Chen}}, \bibinfo {author} {\bibfnamefont {Z.}~\bibnamefont {Yan}}, \bibinfo {author} {\bibfnamefont {G.}~\bibnamefont {Lin}},\ and\ \bibinfo {author} {\bibfnamefont {B.}~\bibnamefont {Andresen}},\ }\bibfield  {title} {\bibinfo {title} {On the curzon--ahlborn efficiency and its connection with the efficiencies of real heat engines},\ }\href@noop {} {\bibfield  {journal} {\bibinfo  {journal} {Energy Convers. Manag.}\ }\textbf {\bibinfo {volume} {42}},\ \bibinfo {pages} {173} (\bibinfo {year} {2001})}\BibitemShut {NoStop}%
\bibitem [{\citenamefont {Whitney}(2014)}]{PhysRevLett.112.130601}%
  \BibitemOpen
  \bibfield  {author} {\bibinfo {author} {\bibfnamefont {R.~S.}\ \bibnamefont {Whitney}},\ }\bibfield  {title} {\bibinfo {title} {Most efficient quantum thermoelectric at finite power output},\ }\href {https://doi.org/10.1103/PhysRevLett.112.130601} {\bibfield  {journal} {\bibinfo  {journal} {Phys. Rev. Lett.}\ }\textbf {\bibinfo {volume} {112}},\ \bibinfo {pages} {130601} (\bibinfo {year} {2014})}\BibitemShut {NoStop}%
\bibitem [{\citenamefont {Whitney}(2015)}]{PhysRevB.91.115425}%
  \BibitemOpen
  \bibfield  {author} {\bibinfo {author} {\bibfnamefont {R.~S.}\ \bibnamefont {Whitney}},\ }\bibfield  {title} {\bibinfo {title} {Finding the quantum thermoelectric with maximal efficiency and minimal entropy production at given power output},\ }\href {https://doi.org/10.1103/PhysRevB.91.115425} {\bibfield  {journal} {\bibinfo  {journal} {Phys. Rev. B}\ }\textbf {\bibinfo {volume} {91}},\ \bibinfo {pages} {115425} (\bibinfo {year} {2015})}\BibitemShut {NoStop}%
\bibitem [{\citenamefont {Ryabov}\ and\ \citenamefont {Holubec}(2016)}]{PhysRevE.93.050101}%
  \BibitemOpen
  \bibfield  {author} {\bibinfo {author} {\bibfnamefont {A.}~\bibnamefont {Ryabov}}\ and\ \bibinfo {author} {\bibfnamefont {V.}~\bibnamefont {Holubec}},\ }\bibfield  {title} {\bibinfo {title} {Maximum efficiency of steady-state heat engines at arbitrary power},\ }\href {https://doi.org/10.1103/PhysRevE.93.050101} {\bibfield  {journal} {\bibinfo  {journal} {Phys. Rev. E}\ }\textbf {\bibinfo {volume} {93}},\ \bibinfo {pages} {050101} (\bibinfo {year} {2016})}\BibitemShut {NoStop}%
\bibitem [{\citenamefont {Raux}\ \emph {et~al.}(2025)\citenamefont {Raux}, \citenamefont {Goupil},\ and\ \citenamefont {Verley}}]{raux2025three}%
  \BibitemOpen
  \bibfield  {author} {\bibinfo {author} {\bibfnamefont {P.}~\bibnamefont {Raux}}, \bibinfo {author} {\bibfnamefont {C.}~\bibnamefont {Goupil}},\ and\ \bibinfo {author} {\bibfnamefont {G.}~\bibnamefont {Verley}},\ }\bibfield  {title} {\bibinfo {title} {Three optima of thermoelectric conversion: Insights from the constant property model},\ }\href@noop {} {\bibfield  {journal} {\bibinfo  {journal} {Entropy}\ }\textbf {\bibinfo {volume} {27}},\ \bibinfo {pages} {252} (\bibinfo {year} {2025})}\BibitemShut {NoStop}%
\bibitem [{\citenamefont {Johal}(2017)}]{PhysRevE.96.012151}%
  \BibitemOpen
  \bibfield  {author} {\bibinfo {author} {\bibfnamefont {R.~S.}\ \bibnamefont {Johal}},\ }\bibfield  {title} {\bibinfo {title} {Heat engines at optimal power: Low-dissipation versus endoreversible model},\ }\href {https://doi.org/10.1103/PhysRevE.96.012151} {\bibfield  {journal} {\bibinfo  {journal} {Phys. Rev. E}\ }\textbf {\bibinfo {volume} {96}},\ \bibinfo {pages} {012151} (\bibinfo {year} {2017})}\BibitemShut {NoStop}%
\bibitem [{\citenamefont {Zhang}\ and\ \citenamefont {Huang}(2020)}]{PhysRevE.102.012151}%
  \BibitemOpen
  \bibfield  {author} {\bibinfo {author} {\bibfnamefont {Y.}~\bibnamefont {Zhang}}\ and\ \bibinfo {author} {\bibfnamefont {Y.}~\bibnamefont {Huang}},\ }\bibfield  {title} {\bibinfo {title} {Applicability of the low-dissipation model: Carnot-like heat engines under newton's law of cooling},\ }\href {https://doi.org/10.1103/PhysRevE.102.012151} {\bibfield  {journal} {\bibinfo  {journal} {Phys. Rev. E}\ }\textbf {\bibinfo {volume} {102}},\ \bibinfo {pages} {012151} (\bibinfo {year} {2020})}\BibitemShut {NoStop}%
\bibitem [{\citenamefont {Gonzalez-Ayala}\ \emph {et~al.}(2017)\citenamefont {Gonzalez-Ayala}, \citenamefont {Roco}, \citenamefont {Medina},\ and\ \citenamefont {Calvo~Hern{\'a}ndez}}]{gonzalez2017carnot}%
  \BibitemOpen
  \bibfield  {author} {\bibinfo {author} {\bibfnamefont {J.}~\bibnamefont {Gonzalez-Ayala}}, \bibinfo {author} {\bibfnamefont {J.~M.~M.}\ \bibnamefont {Roco}}, \bibinfo {author} {\bibfnamefont {A.}~\bibnamefont {Medina}},\ and\ \bibinfo {author} {\bibfnamefont {A.}~\bibnamefont {Calvo~Hern{\'a}ndez}},\ }\bibfield  {title} {\bibinfo {title} {Carnot-like heat engines versus low-dissipation models},\ }\href@noop {} {\bibfield  {journal} {\bibinfo  {journal} {Entropy}\ }\textbf {\bibinfo {volume} {19}},\ \bibinfo {pages} {182} (\bibinfo {year} {2017})}\BibitemShut {NoStop}%
\bibitem [{\citenamefont {Guo}\ \emph {et~al.}(2023)\citenamefont {Guo}, \citenamefont {Ke},\ and\ \citenamefont {Jiang}}]{guo2023performance}%
  \BibitemOpen
  \bibfield  {author} {\bibinfo {author} {\bibfnamefont {J.}~\bibnamefont {Guo}}, \bibinfo {author} {\bibfnamefont {Q.}~\bibnamefont {Ke}},\ and\ \bibinfo {author} {\bibfnamefont {F.}~\bibnamefont {Jiang}},\ }\bibfield  {title} {\bibinfo {title} {The performance of practical refrigerators: A comparative study of the endoreversible and low-dissipation models},\ }\href@noop {} {\bibfield  {journal} {\bibinfo  {journal} {Case Stud. Therm. Eng.}\ }\textbf {\bibinfo {volume} {41}},\ \bibinfo {pages} {102634} (\bibinfo {year} {2023})}\BibitemShut {NoStop}%
\bibitem [{\citenamefont {Gonzalez-Ayala}\ \emph {et~al.}(2016)\citenamefont {Gonzalez-Ayala}, \citenamefont {Hern{\'a}ndez},\ and\ \citenamefont {Roco}}]{gonzalez2016irreversible}%
  \BibitemOpen
  \bibfield  {author} {\bibinfo {author} {\bibfnamefont {J.}~\bibnamefont {Gonzalez-Ayala}}, \bibinfo {author} {\bibfnamefont {A.~C.}\ \bibnamefont {Hern{\'a}ndez}},\ and\ \bibinfo {author} {\bibfnamefont {J.}~\bibnamefont {Roco}},\ }\bibfield  {title} {\bibinfo {title} {Irreversible and endoreversible behaviors of the ld-model for heat devices: the role of the time constraints and symmetries on the performance at maximum $\chi$ figure of merit},\ }\href@noop {} {\bibfield  {journal} {\bibinfo  {journal} {J. Stat. Mech. Theor. Exp.}\ }\textbf {\bibinfo {volume} {2016}},\ \bibinfo {pages} {073202} (\bibinfo {year} {2016})}\BibitemShut {NoStop}%
\bibitem [{\citenamefont {Hern\'andez}\ \emph {et~al.}(2001)\citenamefont {Hern\'andez}, \citenamefont {Medina}, \citenamefont {Roco}, \citenamefont {White},\ and\ \citenamefont {Velasco}}]{PhysRevE.63.037102}%
  \BibitemOpen
  \bibfield  {author} {\bibinfo {author} {\bibfnamefont {A.~C.}\ \bibnamefont {Hern\'andez}}, \bibinfo {author} {\bibfnamefont {A.}~\bibnamefont {Medina}}, \bibinfo {author} {\bibfnamefont {J.~M.~M.}\ \bibnamefont {Roco}}, \bibinfo {author} {\bibfnamefont {J.~A.}\ \bibnamefont {White}},\ and\ \bibinfo {author} {\bibfnamefont {S.}~\bibnamefont {Velasco}},\ }\bibfield  {title} {\bibinfo {title} {Unified optimization criterion for energy converters},\ }\href {https://doi.org/10.1103/PhysRevE.63.037102} {\bibfield  {journal} {\bibinfo  {journal} {Phys. Rev. E}\ }\textbf {\bibinfo {volume} {63}},\ \bibinfo {pages} {037102} (\bibinfo {year} {2001})}\BibitemShut {NoStop}%
\bibitem [{\citenamefont {Angulo-Brown}(1991)}]{angulo1991ecological}%
  \BibitemOpen
  \bibfield  {author} {\bibinfo {author} {\bibfnamefont {F.}~\bibnamefont {Angulo-Brown}},\ }\bibfield  {title} {\bibinfo {title} {An ecological optimization criterion for finite-time heat engines},\ }\href@noop {} {\bibfield  {journal} {\bibinfo  {journal} {J. Appl. Phys.}\ }\textbf {\bibinfo {volume} {69}},\ \bibinfo {pages} {7465} (\bibinfo {year} {1991})}\BibitemShut {NoStop}%
\bibitem [{\citenamefont {S\'anchez-Salas}\ \emph {et~al.}(2010)\citenamefont {S\'anchez-Salas}, \citenamefont {L\'opez-Palacios}, \citenamefont {Velasco},\ and\ \citenamefont {Calvo~Hern\'andez}}]{PhysRevE.82.051101}%
  \BibitemOpen
  \bibfield  {author} {\bibinfo {author} {\bibfnamefont {N.}~\bibnamefont {S\'anchez-Salas}}, \bibinfo {author} {\bibfnamefont {L.}~\bibnamefont {L\'opez-Palacios}}, \bibinfo {author} {\bibfnamefont {S.}~\bibnamefont {Velasco}},\ and\ \bibinfo {author} {\bibfnamefont {A.}~\bibnamefont {Calvo~Hern\'andez}},\ }\bibfield  {title} {\bibinfo {title} {Optimization criteria, bounds, and efficiencies of heat engines},\ }\href {https://doi.org/10.1103/PhysRevE.82.051101} {\bibfield  {journal} {\bibinfo  {journal} {Phys. Rev. E}\ }\textbf {\bibinfo {volume} {82}},\ \bibinfo {pages} {051101} (\bibinfo {year} {2010})}\BibitemShut {NoStop}%
\bibitem [{\citenamefont {Zhang}\ \emph {et~al.}(2016)\citenamefont {Zhang}, \citenamefont {Huang}, \citenamefont {Lin},\ and\ \citenamefont {Chen}}]{PhysRevE.93.032152}%
  \BibitemOpen
  \bibfield  {author} {\bibinfo {author} {\bibfnamefont {Y.}~\bibnamefont {Zhang}}, \bibinfo {author} {\bibfnamefont {C.}~\bibnamefont {Huang}}, \bibinfo {author} {\bibfnamefont {G.}~\bibnamefont {Lin}},\ and\ \bibinfo {author} {\bibfnamefont {J.}~\bibnamefont {Chen}},\ }\bibfield  {title} {\bibinfo {title} {Universality of efficiency at unified trade-off optimization},\ }\href {https://doi.org/10.1103/PhysRevE.93.032152} {\bibfield  {journal} {\bibinfo  {journal} {Phys. Rev. E}\ }\textbf {\bibinfo {volume} {93}},\ \bibinfo {pages} {032152} (\bibinfo {year} {2016})}\BibitemShut {NoStop}%
\bibitem [{\citenamefont {Apertet}\ \emph {et~al.}(2013)\citenamefont {Apertet}, \citenamefont {Ouerdane}, \citenamefont {Michot}, \citenamefont {Goupil},\ and\ \citenamefont {Lecoeur}}]{apertet2013efficiency}%
  \BibitemOpen
  \bibfield  {author} {\bibinfo {author} {\bibfnamefont {Y.}~\bibnamefont {Apertet}}, \bibinfo {author} {\bibfnamefont {H.}~\bibnamefont {Ouerdane}}, \bibinfo {author} {\bibfnamefont {A.}~\bibnamefont {Michot}}, \bibinfo {author} {\bibfnamefont {C.}~\bibnamefont {Goupil}},\ and\ \bibinfo {author} {\bibfnamefont {P.}~\bibnamefont {Lecoeur}},\ }\bibfield  {title} {\bibinfo {title} {On the efficiency at maximum cooling power},\ }\href@noop {} {\bibfield  {journal} {\bibinfo  {journal} {EPL}\ }\textbf {\bibinfo {volume} {103}},\ \bibinfo {pages} {40001} (\bibinfo {year} {2013})}\BibitemShut {NoStop}%
\bibitem [{\citenamefont {Guo}\ \emph {et~al.}(2020)\citenamefont {Guo}, \citenamefont {Yang}, \citenamefont {Gonzalez-Ayala}, \citenamefont {Roco}, \citenamefont {Medina},\ and\ \citenamefont {Hern{\'a}ndez}}]{guo2020equivalent}%
  \BibitemOpen
  \bibfield  {author} {\bibinfo {author} {\bibfnamefont {J.}~\bibnamefont {Guo}}, \bibinfo {author} {\bibfnamefont {H.}~\bibnamefont {Yang}}, \bibinfo {author} {\bibfnamefont {J.}~\bibnamefont {Gonzalez-Ayala}}, \bibinfo {author} {\bibfnamefont {J.}~\bibnamefont {Roco}}, \bibinfo {author} {\bibfnamefont {A.}~\bibnamefont {Medina}},\ and\ \bibinfo {author} {\bibfnamefont {A.~C.}\ \bibnamefont {Hern{\'a}ndez}},\ }\bibfield  {title} {\bibinfo {title} {The equivalent low-dissipation combined cycle system and optimal analyses of a class of thermally driven heat pumps},\ }\href@noop {} {\bibfield  {journal} {\bibinfo  {journal} {Energy Convers. Manag.}\ }\textbf {\bibinfo {volume} {220}},\ \bibinfo {pages} {113100} (\bibinfo {year} {2020})}\BibitemShut {NoStop}%
\end{thebibliography}%

\end{document}